\documentclass[acmsmall]{acmart}

\AtBeginDocument{%
  \providecommand\BibTeX{{%
    \normalfont B\kern-0.5em{\scshape i\kern-0.25em b}\kern-0.8em\TeX}}}

\usepackage{caption}
\usepackage{subcaption}
\usepackage{booktabs}
\usepackage{multirow}
\usepackage{algorithm}
\usepackage{algpseudocode}
\usepackage{soul}
\usepackage{color, colortbl}
\definecolor{Gray}{gray}{0.9}
\usepackage[utf8]{inputenc}
\acmJournal{TQC}
\acmVolume{}
\acmNumber{}
\acmArticle{}
\acmMonth{8}

\begin{document}

\title{Robust Qubit Mapping Algorithm via Double-Source Optimal Routing on Large Quantum Circuits}


% \IEEEauthorblockA{\IEEEauthorrefmark{1}Department of Electrical Engineering, National Taiwan University, Taiwan}
% \IEEEauthorblockA{\IEEEauthorrefmark{2}Graduate Institute of Communication Engineering, National Taiwan University, Taiwan}

% Needed for Hao-Chung
%\IEEEauthorblockA{\IEEEauthorrefmark{3}Center for Quantum Science and Engineering,  National Taiwan University, Taiwan}

% Needed for Hao-Chung \IEEEauthorblockA{\IEEEauthorrefmark{4}Physics Division, National Center for Theoretical Sciences,  Taiwan}

% Needed for Hao-Chung
%\IEEEauthorblockA{\IEEEauthorrefmark{5}Hon Hai (Foxconn) Quantum Computing Center,  Taiwan}
% \IEEEauthorblockA{\IEEEauthorrefmark{6}Department of Computer Science and Engineering, University of California San Diego}
% \IEEEauthorblockA{\IEEEauthorrefmark{7}College of Computing, The Georgia Institute of Technology, USA}% <-this % stops an unwanted space
% james31423a@gmail.com, chy036@ucsd.edu, rentruewang@gatech.edu, b08901173@ntu.edu.tw, \\haochung.ch@gmail.com, cyhuang@ntu.edu.tw
% }

\author{Chin-Yi Cheng}
\affiliation{%
  \institution{Department of Electrical Engineering, National Taiwan University}
  \country{Taiwan~(R.O.C.)}
  }
\email{r10921132@ntu.edu.tw}

\author{Chien-Yi Yang}
\affiliation{%
  \institution{Department of Computer Science and Engineering, University of California San Diego}
  \country{USA}
  }
\email{chy036@ucsd.edu}

\author{Yi-Hsiang Kuo}
\affiliation{%
  \institution{Graduate Institute of Electronics Engineering, National Taiwan University}
  \country{Taiwan~(R.O.C.)}
}
\email{r12943104@ntu.edu.tw}

\author{Ren-Chu Wang}
\affiliation{%
  \institution{College of Computing, The Georgia Institute of Technology}
  % \city{Rocquencourt}
  \country{USA}
}
\email{rentruewang@gatech.edu}

\author{Hao-Chung Cheng}
\affiliation{%
  \institution{Department of Electrical Engineering
  and Graduate Institute of Communication Engineering, National Taiwan University}
  % \city{Rocquencourt}
  \country{Taiwan (R.O.C.)}
}
\additionalaffiliation{
\institution{Center for Quantum Science and Engineering, National Taiwan University}
% \department{Center for Quantum Science and Engineering}
  % \city{Rocquencourt}
  \country{Taiwan (R.O.C.)}
}
\additionalaffiliation{
\institution{Physics Division and Mathematics Division, National Center for Theoretical Sciences}
% \department{Physics Division}
  % \city{Rocquencourt}
  \country{Taiwan (R.O.C.)}
}
\additionalaffiliation{
\institution{Hon Hai (Foxconn) Quantum Computing Center}
  % \city{Rocquencourt}
  \country{Taiwan (R.O.C.)}
}
\email{haochung.ch@gmail.com}

% Needed for Hao-Chung
%\IEEEauthorblockA{\IEEEauthorrefmark{3}Center for Quantum Science and Engineering,  National Taiwan University, Taiwan}

% Needed for Hao-Chung \IEEEauthorblockA{\IEEEauthorrefmark{4}Physics Division, National Center for Theoretical Sciences,  Taiwan}

% Needed for Hao-Chung
%\IEEEauthorblockA{\IEEEauthorrefmark{5}Hon Hai (Foxconn) Quantum Computing Center,  Taiwan}

\author{Chung-Yang (Ric) Huang}
\affiliation{%
  \institution{Graduate Institute of Electronics Engineering, National Taiwan University}
  % \city{Rocquencourt}
  \country{Taiwan (R.O.C.)}
}
\additionalaffiliation{
\institution{Department of Electrical Engineering, National Taiwan University}
% \department{Center for Quantum Science and Engineering}
  % \city{Rocquencourt}
  \country{Taiwan (R.O.C.)}
}
\email{cyhuang@ntu.edu.tw}


\begin{abstract}
 Qubit Mapping is a critical aspect of implementing quantum circuits on real hardware devices. Currently, the existing algorithms for qubit mapping encounter difficulties when dealing with larger circuit sizes involving hundreds of qubits. In this paper, we introduce an innovative qubit mapping algorithm, Duostra, tailored to address the challenge of implementing large-scale quantum circuits on real hardware devices with limited connectivity. Duostra operates by efficiently determining optimal paths for double-qubit gates and inserting SWAP gates accordingly to implement the double-qubit operations on real devices. Together with two heuristic scheduling algorithms, the Limitedly-Exhausitive (LE) Search and the Shortest-Path (SP) Estimation, it yields results of good quality within a reasonable runtime, thereby striving toward achieving quantum advantage. Experimental results showcase our algorithm's superiority, especially for large circuits beyond the NISQ era. {\color{black}For example, on large circuits with more than 50 qubits, we can reduce the mapping cost on an average 21.75\% over the virtual best results among QMAP, t$|ket\rangle$, Qiskit and SABRE. Besides, for mid-size circuits such as the SABRE-large benchmark, we improve the mapping costs by 4.5\%, 5.2\%, 16.3\%, 20.7\%, and 25.7\%, when compared to QMAP, TOQM, t$|ket\rangle$, Qiskit, and SABRE, respectively.}
\end{abstract}

% <concept_id>10010520.10010521.10010542.10010550</concept_id>
% <concept_desc>Computer systems organization~Quantum computing</concept_desc>
% <concept_id>10010583.10010682.10010697.10010702</concept_id>
% <concept_desc>Hardware~Physical synthesis</concept_desc>
% <concept_significance>300</concept_significance>

\ccsdesc[500]{Computer systems organization~Quantum computing}
\ccsdesc[300]{Hardware~Physical synthesis}

\keywords{Quantum Compilation, Qubit Mapping, Scalability, Local Optimal Solution}


\maketitle

\section{Introduction}
The ultimate goal of quantum computing is to outperform the classical algorithms exponentially in solving real-world intractable problems \cite{nielsen_quantum_2010}, including Quantum Communication \cite{khatri_principles_2020}, Quantum Cryptography \cite{chen_quantum_2015}, and Quantum Machine Learning \cite{arunachalam_survey_2017}, etc. Recent progress in quantum computing devices has sparked hope for this dream to become a reality. For instance, IBM Condor has reached a milestone of 1,121 qubits, tripling its previous generation, Osprey, from the previous year. This is only about $1/2$ of the 2048 qubits needed for the Quantum Fourier Transform (QFT) to break the RSA-1024 encryption \cite{shor_algorithms_1994,shor_polynomial-time_1997}, and approximately $1/30$ of the $255\times6$ qubits required for the discrete logarithms over elliptic curves (e.g., Curve25519) to solve the Bitcoin encryption \cite{proos_shors_2003}. 

\begin{figure}[!t]
\centering
\includegraphics[scale=0.25]{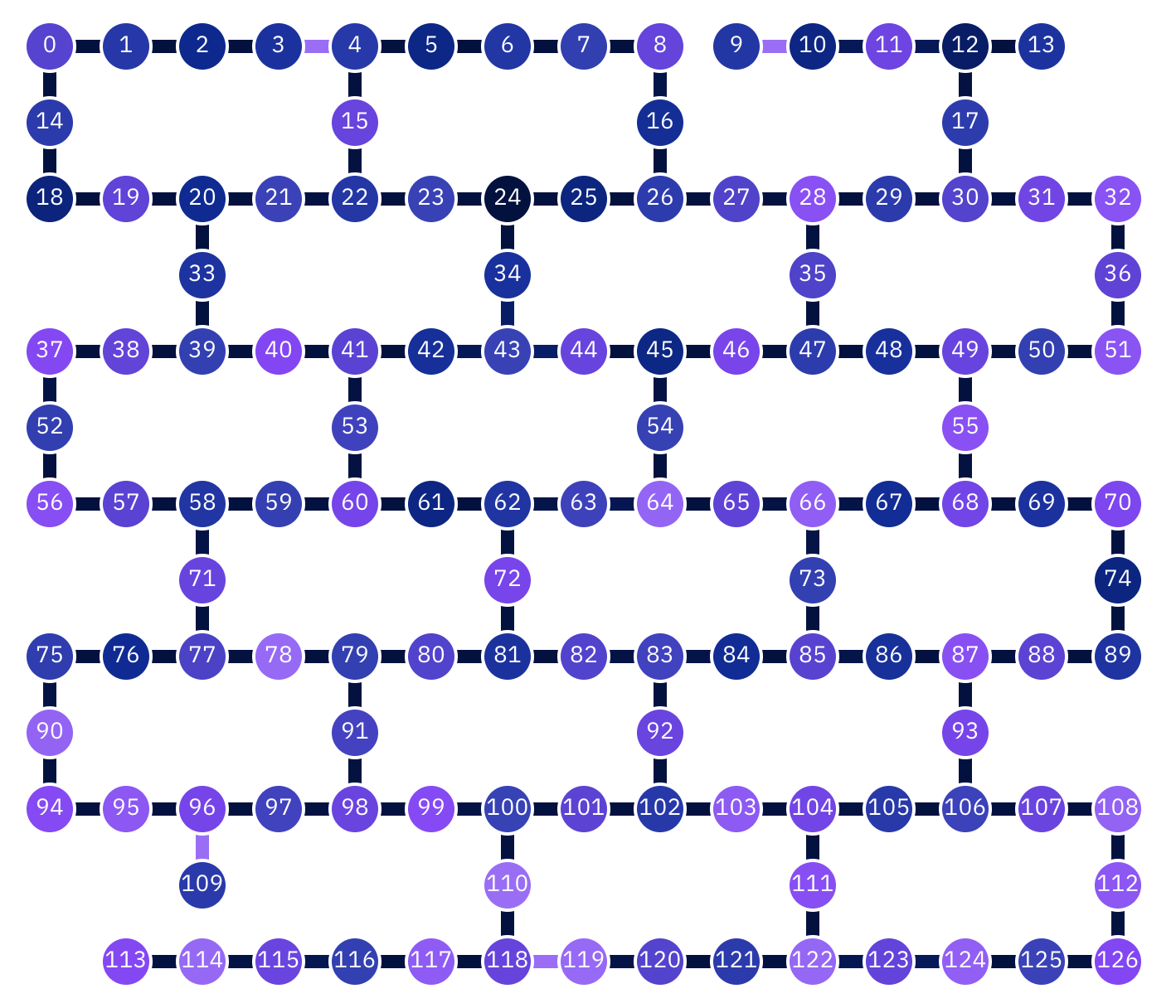}
\caption{The topology of \emph{ibmq\_washington}, retrieved from \cite{ibm_ibm_2021}}
\label{fig:IBMtopo}
\end{figure}

However, the dream of achieving the quantum advantage can be seriously deferred by the  ``connectivity constraints'' of modern quantum computing devices. Taking the \emph{ibmq\_washington} quantum computer of which the topology is shown in Fig.~\ref{fig:IBMtopo} as an example, each of its qubits (i.e. a vertex in the graph) is connected to at most three adjacent qubits. Therefore, when a quantum algorithm involves a quantum operation between two non-adjacent qubits, say qubits 0 and 126 at the extreme, we will need a long sequence of swap operations to bring the computation to a pair of adjacent qubits, say qubits 61 and 62. This will significantly increase the demand for the number of qubits and, even worse, lead to a much longer execution time for the quantum circuit. Therefore, intelligently mapping the quantum circuit to the physical qubits, known as the ``qubit mapping problem,'' will be one of the most crucial steps in achieving the quantum advantage.

The qubit mapping problem involves an initial placement of the logical qubits to the physical ones, and subsequent SWAP insertions to bring the computations of multiple qubit gates to adjacent qubits. {\color{black}The objective is to improve the circuit fidelity. Common approaches include reducing the number of SWAPs and thus optimizing the execution time of the quantum circuit and mitigating the error of each gate and qubit. }Siraichi \emph{et al.} \cite{siraichi_qubit_2018} have shown qubit mapping to be an NP-complete problem. 

Earlier algorithms  \cite{maslov_quantum_2008,saeedi_synthesis_2011,chakrabarti_linear_2011,shafaei_optimization_2013,shafaei_qubit_2014,wille_optimal_2014,wille_exact_2014,lye_determining_2015,pedram_layout_2016,wille_look-ahead_2016} attempted to minimize the number of additional SWAP gates with the mapping of qubits to the Linear Nearest Neighbor (LNN) structure. Booth \emph{et al.}~\cite{booth_comparing_2018} even transformed the mapping into a mathematical problem and utilized solvers to obtain feasible results. However, modern quantum computing devices in the NISQ era are mostly in lattice structures, which offer higher routing flexibility than the LNN structure. Therefore, the aforementioned approaches could only achieve sub-optimal solutions for the qubit mapping problem.

To realize the qubit mapping on the NISQ devices, Siraichi \textit{et al.} \cite{siraichi_qubit_2018} introduced a straightforward heuristic approach that worked on the lattice structure of the earlier generations of the IBM quantum computers. {\color{black}de Almeida \textit{et al.} \cite{de_almeida_cnot_2019} relied on the matrix cost to solve CNOT mappings on IBM QX device.} However, as the qubit interaction of these early-day machines is unidirectional, their algorithms are not suitable for modern quantum devices that pose bidirectional qubit interactions. {\color{black}There were studies focusing on reducing depth overheads besides the number of SWAP gates. Venturelli \textit{et al.} \cite{venturelli_compiling_2018} used exhaustive-search temporal planners to schedule gates.} Zulehner \textit{et al.} \cite{zulehner_efficient_2018,zulehner_compiling_2019} and {\color{black}Matsuo \emph{et al.} \cite{matsuo_reducing_2019}} then proposed a multi-step approach with a depth-based layer partitioning and A* as the underlying search algorithm. Additionally, Peham \textit{et al.} \cite{peham_optimal_2022} explored the sub-architectures of the quantum devices in order to devise a suitable initial placement strategy for the qubit mapping problem. Notably, \cite{zulehner_efficient_2018} and \cite{peham_optimal_2022} have been integrated into the QMAP framework \cite{wille_mqt_2023}, which delivered high-quality results for relatively small quantum circuits (less than 50 qubits). However, when applied to larger circuits, QMAP will suffer in lengthier program runtime and thus the mapping outcomes are not as effective. Li \textit{et al.} \cite{li_tackling_2019} introduced SABRE\footnote{\textbf{S}W\textbf{A}P-based \textbf{B}idi\textbf{RE}ctional heuristic search algorithm}, employing a heuristic based on shortest path costs and look-ahead techniques to minimize the circuit depth of the quantum circuits. {\color{black} Zhu \emph{et al.} \cite{zhu_dynamic_2020} and Jiang \emph{et al.} \cite{jiang_quantum_2021} improved SABRE by lookahead methods. In addition to SABRE-inspired research, Guerreschi \emph{et al.} \cite{guerreschi_two-step_2018} also introduced a technique aimed at minimizing the overhead of inserted SWAPs. This method involved initially organizing quantum operations based on logical dependencies, followed by rescheduling according to physical constraints. Nonetheless, when compared to the subsequent advancements, their program runtime and/or the quality of the solution are inferior.}

To counter the solution quality problems of the above-mentioned heuristic-based methods, several researchers proposed methodologies that rendered the exact solution for the optimal result with respect to the number of inserted SWAPs. In addition to the straightforward heuristic approach, Siraichi \emph{et al.} \cite{siraichi_qubit_2018} also presented an exhaustive computation based on the dynamic programming algorithm. However, it comes with the complexity of $O((Q!)^2)$, where $Q$ is the number of qubits, and thus is not applicable for larger quantum circuits. Some other exact-solution approaches formulated the qubit mapping problem into a satisfiability problem. Wille et al. \textit {et al.} \cite{wille_exact_2014, wille_mapping_2019} utilized the Satisfiability Modulo Theories (SMT) solver together with the pseudo-Boolean optimizer to minimize the inserted SWAP gates. Tan and Cong \cite{tan_optimal_2020} constructed a gate-dependency graph and guaranteed optimality for objectives either in circuit depths or SWAP counts via an SMT solver named OLSQ. Their follow-up works, namely OLSQ-GA \cite{tan_optimal_2021} and OLSQ2 \cite{lin_scalable_2023}, claimed to enlarge the scalability to 54 qubits. {\color{black}Moreover, Molavi \textit{et al.} \cite{molavi_qubit_2022} utilized maximum satisfiability (MaxSAT) to minimize the mapping cost.} However, these exact methods, although being able to achieve optimality for small circuits, are inferior to the heuristic methods in large circuits. It is often that they will abort on large circuits due to the complex exhausting search process.

{\color{black} Another technique to improve the solution quality of the qubit mapping is to define the timing model for distinct gate types so that the mapping cost can better approximate the execution time of the quantum circuit on the real device. Consequently, the optimal qubit mapping solution can ensure the better fidelity of the quantum computation.} Deng~\emph{et al.}~\cite{deng_codar_2020} proposed a timing model for super-conducting devices that defined the timing costs of the single-qubit, double-qubit, and SWAP gates to be 1, 2, and 6 units, respectively. The mapping cost is therefore calculated based on these timing costs. They further introduced the concept of lock time for the busy qubits and presented it as the CODAR\footnote{\textbf{CO}ntext-sensitive and \textbf{D}uration-\textbf{A}ware \textbf{R}emapping algorithm} algorithm to optimize the total circuit execution time. On the other hand, Zhang \emph{et al.} in \cite{zhang_time-optimal_2021} adopted a similar timing model\footnote{1, 1, and 3 units for the single-qubit, double-qubit, and SWAP gates} and devised an improved algorithm, Time-Optimal Qubit Mapping (TOQM), with estimated cost to achieve better results for the mapping. {\color{black} Lao et al. \cite{lao_timing_2022} proposed a Qmap\footnote{Note: Qmap \cite{lao_timing_2022} and QMAP \cite{wille_mqt_2023} are two different studies.} algorithm for Surface-17, a 17-qubit quantum computer. To accommodate the specific controlling mechanism of this device, they imposed frequency constraints into mapping costs and minimized the execution time by exhaustive search on the potential routings. However, their runtime complexity can be as high as $O(g\sqrt{n}4^{\sqrt{n}})$, where $g$ and $n$ are numbers of gates and qubits respectively. Although the above-mentioned algorithms claimed to achieve optimal mapping solutions, the runtime of these algorithms was very long due to the fact that they needed to calculate the costs of all the potential swapping candidates in each iteration when selecting a SWAP. Consequently, they could only handle mapping problems up to 50 qubits, which is much smaller than the number of required qubits for achieving quantum advantage.}  

{\color{black}This paper introduces a robust qubit mapping framework designed for large circuits with hundreds of qubits and beyond. Our key contributions are two-fold:} 
\begin{enumerate}
    \item {\color{black}While many previous approaches such as \cite{deng_codar_2020}, \cite{zhang_time-optimal_2021}, and \cite{lao_timing_2022} have exponential or sub-exponential complexities in finding the optimal SWAP insertion sequence, we propose a qubit mapping framework that decomposes the optimization flow into three components: an $O(n)$ initial mapping strategy, an $O(n\log{n})$ Duostra router that can guarantee the optimal mapping solution for a given double-qubit gate, and a heuristic scheduler that can either exploit a limited-depth exhaustive search or a shortest-path estimation. The overall algorithm strikes a good balance that it can obtain the near-optimal solution for smaller circuits as well as scale up to handle very large-scale circuits.}
    \item {\color{black}Unlike other previous algorithms that optimize the mapping solutions by considering static costs derived from the device topologies (e.g. SABRE), our proposed Duostra algorithm dynamically computes the ``occupied time'' (defined in section~\ref{section:terminologies}) during the finding of the routing paths. It can not only ensure the qubit mapping optimality for a double-qubit gate, but also update the coupling constraints for the other double-qubit gates so that the scheduler can further optimize the mapping for the subsequent routing paths.}
\end{enumerate}
				
{\color{black}The experimental results indicate that in the mid-size test cases within the SABRE-large test suite \cite{li_tackling_2019}, our approach achieves an average cost improvement by 4.5\%, 5.2\%, 16.3\%, 20.7\%, and 25.7\%, when compared to QMAP, TOQM, t$|ket\rangle$, Qiskit, and SABRE, respectively. Additionally, those results are generated within a runtime of 5 minutes. Regarding scalability, Duostra surpasses IBM Qiskit, QMAP, t$|ket\rangle$, and SABRE on mapping cost for most large quantum circuits, including Galois Field (GF), inst, classical Electronic Design Automation (EDA) benchmark circuits \cite{brglez_neutral_1985}, adder, and qugan. Our framework exhibits an average improvement of 21.75\% over the best results among IBM Qiskit, QMAP, t$|ket\rangle$, and SABRE across 39 large circuits (i.e. qubits > 50), with nearly the shortest runtime. Furthermore, our method can handle problem sizes up to 11,969-qubit QFT within 3 hours (with time complexity $O(n^{2.8})$, where $n$ denotes the numbers of qubits). In summary, our proposed qubit mapping framework offers a solution for large-scale quantum circuits by efficiently implementing local-optimal quantum algorithms on a non-fully connected quantum device.}

The rest of the paper is organized as follows. We first explain the qubit mapping problem in Section \ref{section:QUBITMAPPING}. The proposed framework is detailed in Section \ref{section:FRAMEWORK}, and specifically the proof for the optimality of the Duostra algorithm is provided in Section \ref{section:ProveDuostra}. We present the experiment results in Section \ref{section:EXPERIMENT}, and conclude the paper in Section \ref{section:CONCLUSION}. The source code\footnote{https://github.com/DVLab-NTU/qsyn} of our method is embedded in the tool Qsyn \cite{Lau_Qsyn_2024}.

\section{Qubit Mapping Problem}\label{section:QUBITMAPPING}

The qubit mapping problem is to schedule the operations of {\color{black} a quantum circuit and then assign them to the qubits of a quantum computer with a given topology (e.g. Fig.~\ref{fig:IBMtopo}).} The circuit is composed of gates from a quantum cell library (e.g. \textit{Clifford+T} library). However, we need to insert additional SWAP gates to bring the inputs, which are logically connected but physically apart, of a double-qubit\footnote{Without loss of generality, we assume there are only single- and double-qubit gates in the quantum cell library.} gate to the adjacent qubits. This is because the computation of a double-qubit gate can only be applied on adjacent qubits on the actual hardware device (called ``coupling constraint''), and the topology of the physical device often limits its qubits to interact with only a small number of adjacent neighbors (e.g. $1$ to $3$ for \emph{ibmq\_washington} quantum computer as shown in Fig.~\ref{fig:IBMtopo}).

\begin{figure}
     \begin{minipage}{0.5\textwidth}
     \centering
         \begin{subfigure}[b]{\textwidth}
         \includegraphics[width=\textwidth]{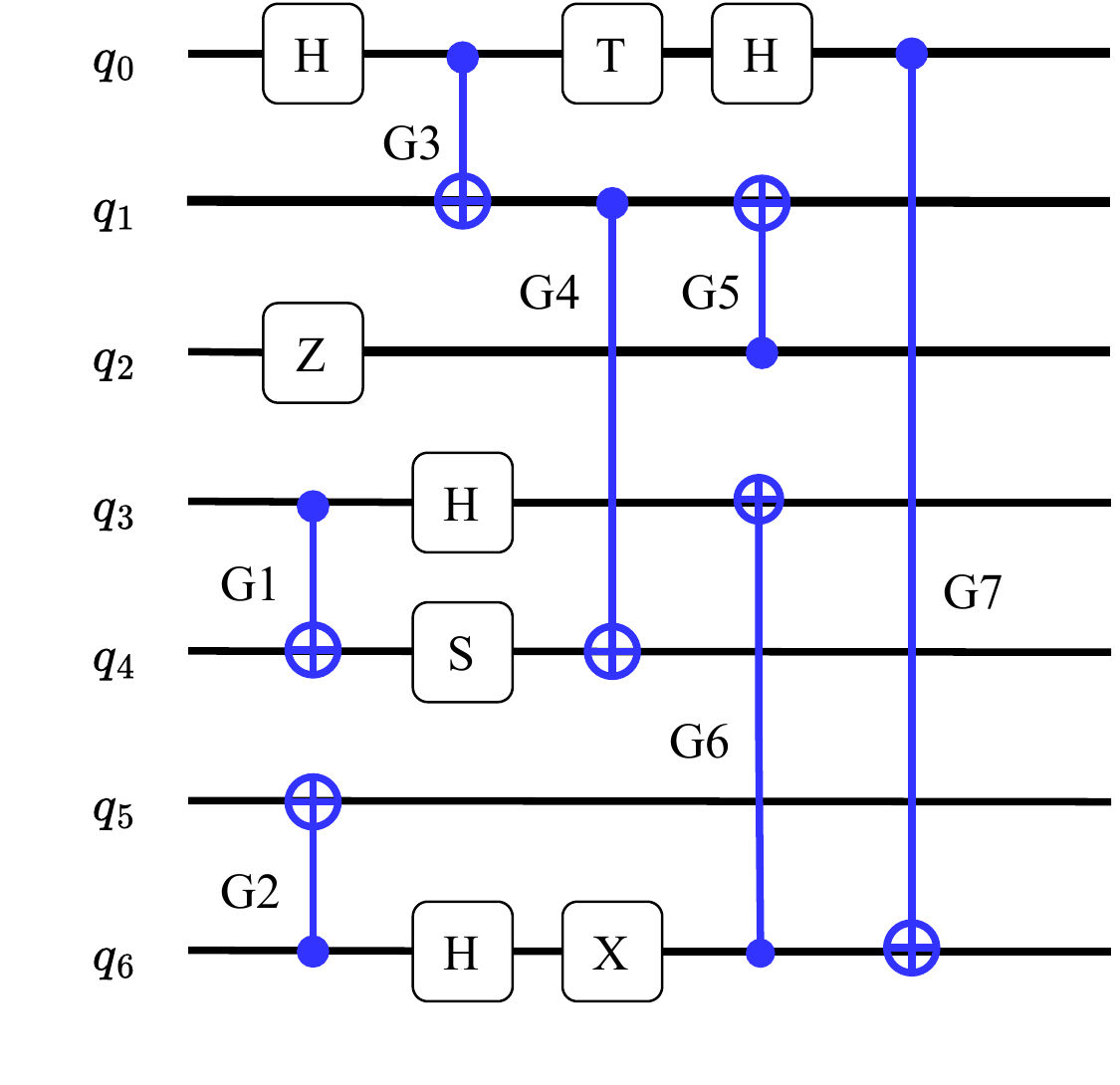}
        \caption{{\color{black}An example of a logical circuit}}
        \label{fig:Dependency Graph-a} 
     \end{subfigure}
     \end{minipage}
     \hfill
     \begin{minipage}{0.4\textwidth}
         \begin{subfigure}[b]{\textwidth}
         \centering
         \includegraphics[width=0.6\textwidth]{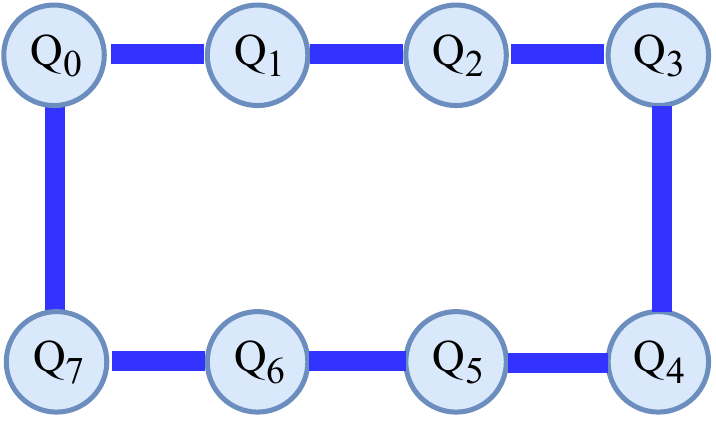}
        \caption{{\color{black}An example of a topology}}
        \label{fig:ring layout} 
     \end{subfigure}
     \hfill
     
     \begin{subfigure}[b]{\textwidth}
         \centering
         \includegraphics[width=0.8\textwidth]{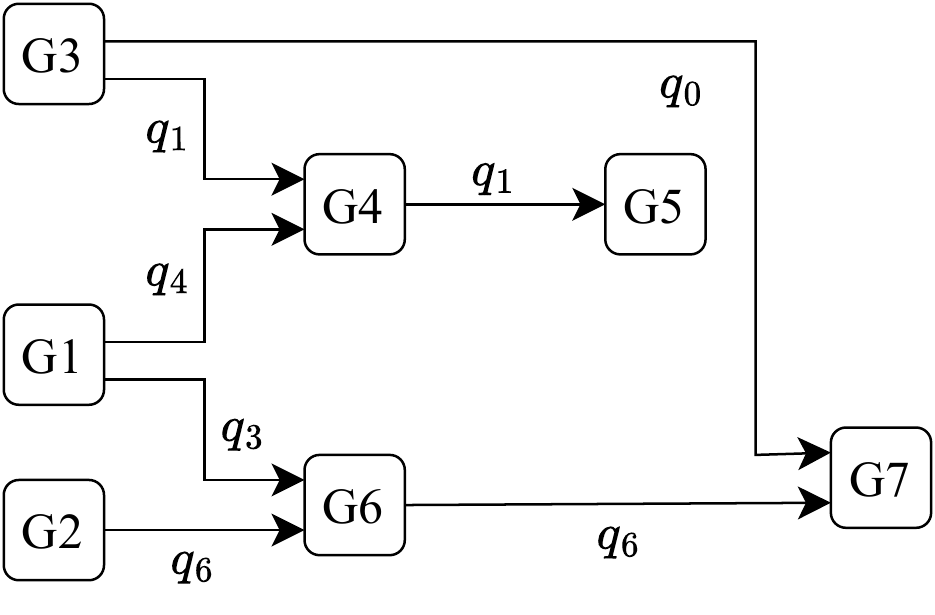}
        \caption{{\color{black}Its corresponding Dependency Graph}}
        \label{fig:Dependency Graph-b} 
     \end{subfigure}
     \end{minipage}
     
     \caption{\color{black}{An example of the qubit mapping problem}}
     \label{fig:Dependency Graph}
\end{figure}

{\color{black} Taking Fig.~\ref{fig:Dependency Graph} as an example of the qubit mapping problem, we denote the logical qubits (i.e. the qubits for a quantum circuit) in lowercase as $q_0$ to $q_6$ in Fig.~\ref{fig:Dependency Graph-a}, and the physical qubits (i.e. the qubits on the actual device) in uppercase as $Q_0$ to $Q_7$ in Fig.~\ref{fig:ring layout}. Let’s assume that after the initial mapping, the inputs of the double-qubit gate $G5$, that is, $q_1$ and $q_2$, are mapped to two non-adjacent physical qubits, $Q_2$ and $Q_6$ (explained later in Section 3.1 for details). We can insert SWAPs along a path, for example, {$Q_2, Q_3, Q_4, Q_5, Q_6$}, to bring these two inputs to adjacency. Clearly, there is another path {$Q_2, Q_1, Q_0, Q_7, Q_6$} to resolve the coupling constraint of $G5$. By choosing one of these two paths, we will employ different numbers of SWAPs and thus relocate the inputs of the subsequent double-qubit gates $G6$ and $G7$ to different physical qubits. This will lead to different execution time of the circuit and therefore different error probabilities.}

A SWAP is composed of three consecutive CX gates, where the middle CX possesses opposite control and target qubits with respect to the other two CX gates. The insertion of the SWAP gates greatly increases the gate counts and execution time of the quantum circuit. Therefore, the objective of the qubit mapping problem is to optimize the scheduling and mapping of the logical cells to the physical qubits and thus minimize the gate counts, execution time, and the resulting computing errors.

To illustrate our qubit mapping framework in the next section, we first introduce four definitions to facilitate the design of the qubit mapping algorithm:

\begin{definition}[Dependency Graph (DepG)]
    A graph that describes the computational dependency of the double-qubit gates in the quantum circuit. For the qubit mapping problem, since the single-qubit gates will never violate the coupling constraints, we only need to consider the double-qubit gates for SWAP insertions to bring their input qubits to the adjacencies. In a Dependency Graph, an edge $Gate_i \rightarrow Gate_j$ means that $Gate_j$ depends on $Gate_i$. A gate can be executed only after all of its parents are executed.
\end{definition}

{\color{black}Take Fig.~\ref{fig:Dependency Graph-a} as an example, the gate $G4$ operates on two qubits, $q_1$ and $q_2$. It can only be executed after all the previous gates (i.e. $G1$ and $G3$) are executed. We say $G4$ depends on $G1$ and $G3$, and add edges from $G1$ and $G3$ to $G4$ to build a directed acyclic graph (DAG),} called Dependency Graph as shown in Fig.~\ref{fig:Dependency Graph-b}.

\begin{definition}[Device Graph (DevG)]
    An undirected graph $DevG = (V, E)$ represents the topology of a real quantum computing device, where $V$ denotes the physical qubits and $E$ is the set of edges corresponding to the connectivity among physical qubits. A pair of qubits can be involved in the double-qubit operation only if there is a direct connection between these two qubits. In the qubit mapping problem, if a double-qubit operation is mapped to two non-adjacent physical qubits, we need to find a path between these two qubits on DevG to perform SWAPs.
\end{definition}

\begin{definition}[Waitlist]

{\color{black} Similar to} the ``front layer'' concept in \cite{li_tackling_2019}, Waitlist is a list of gates that are ready for execution. That is, gates in a Waitlist are those without any unexecuted parents in the Dependency Graph.    
\end{definition}

For example, in Fig.~\ref{fig:Dependency Graph}, {\color{black}if $G2$ and $G3$ were executed, then $G1$ is the only gate in the Waitlist.} Since gates should be executed following the Dependency Graph, gates in the Waitlist are the only candidates when scheduling for the next execution.

\begin{definition}[Ideal Circuit Cost and Mapping Cost]\label{def:cost}
CODAR \cite{deng_codar_2020} notes that the {\color{black} execution time of single-qubit and double-qubit gates is different}. To simplify calculations, Deng \emph{et al.} introduce a model in which single-qubit gates occupy one unit of time and double-qubit gates occupy two units. Using this model, we compute two metrics: Ideal Circuit Cost and Mapping Cost. The Ideal Circuit Cost is derived from the logical circuit before SWAP operations, providing an indication of the circuit's inherent complexity and its performance on a fully connected quantum device. In contrast, Mapping Cost is calculated based on the physical circuit with SWAP insertions taking up six units each. Mapping Cost reflects the time needed for quantum devices to execute the physical circuit. Our subsequent research aims to minimize the Mapping Cost.
\end{definition}

\section{THE PROPOSED QUBIT MAPPING FRAMEWORK AND ALGORITHM}\label{section:FRAMEWORK}

\begin{figure}
    \centering
    \includegraphics[scale=0.5]{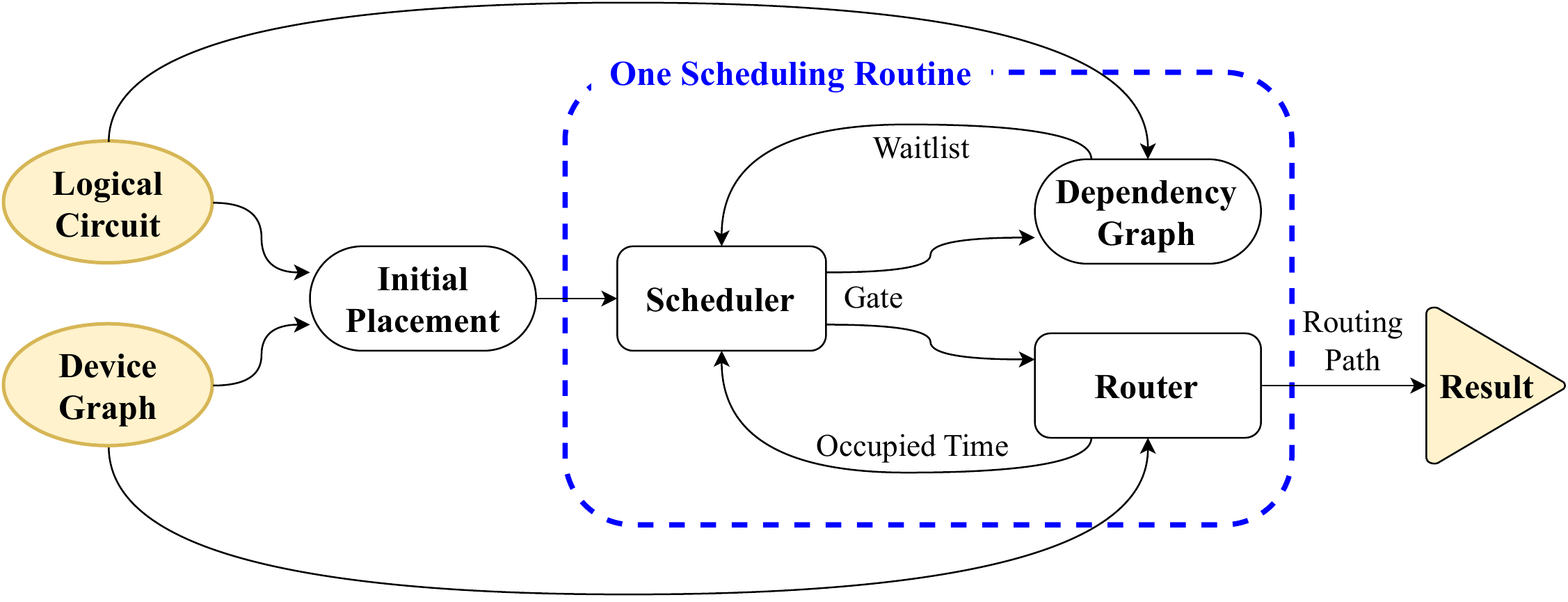}
    \caption{Our qubit mapping framework}
    \label{fig:framework}
\end{figure}

Our proposed qubit mapping framework is shown in Fig.~\ref{fig:framework}. Unlike previous studies which treat the qubit mapping problem as a whole, we propose a framework that decomposes the problem into two manageable subproblems – routing, and scheduling. We start by quickly placing the circuit inputs to certain physical qubits, called initial placement, and from there, we have the first set of gates in the Waitlist. While the Router can efficiently suggest a routing path that minimizes the execution time of a double-qubit gate on a real device, the Scheduler now only needs to consider the order of the gates for execution in the Waitlist. This framework offers flexibility in choosing specific scheduling and routing heuristics with different metrics for different quantum circuits. Below we present our design of algorithms.

\subsection{Initial Placement}\label{section:initialPlacement}

The initial placement step assigns the circuit inputs to specific physical qubits. {\color{black} In our framework, we employ a depth-first search (DFS) strategy for the initial placement. }The primary objective of the DFS placement is to ensure that the two input qubits required by each gate are situated in close proximity to each other.  

\begin{figure}[h]
\centering
    \subfloat[][{\color{black} Adjacency lists of the circuit in Fig.~\ref{fig:Dependency Graph-a}}]{
        \begin{tabular}{c|ccc}
            \toprule
            qubit&\multicolumn{3}{c}{adjacency list}\\ \midrule
            $q_0$ & $q_1$ & $q_6$ & \\
            $q_1$ & $q_0$ & $q_4$ & $q_2$ \\
            $q_2$ & $q_1$ \\
            $q_3$ & $q_4$ & $q_6$ \\
            $q_4$ & $q_3$ & $q_1$\\
            $q_5$ & $q_6$ \\ 
            $q_6$ & $q_5$ & $q_3$ & $q_0$ \\
            \bottomrule
            \end{tabular}
            \vspace{5pt}
            \label{fig: adjacency list}
    }
    \hspace{3cm}
    \subfloat[][{\color{black}Initial placement result}]{
        \begin{subfigure}[b]{0.4\textwidth}
         \centering
        \includegraphics[width=\textwidth]{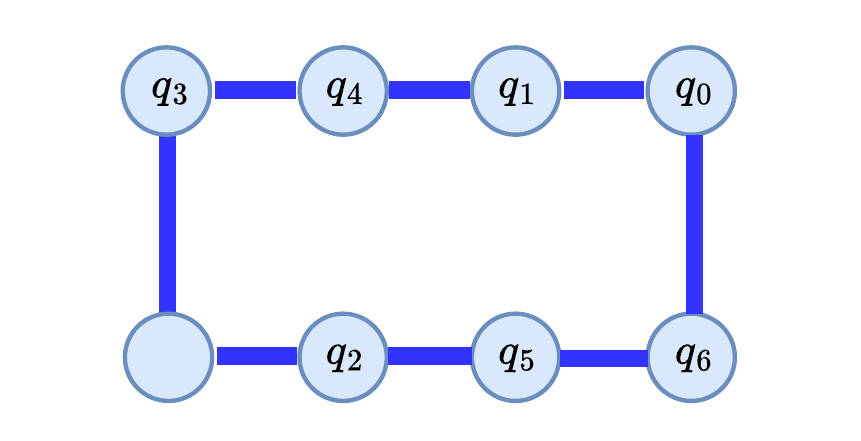}
    \label{fig:placement result}
    \end{subfigure}
    }
    
    \caption{{\color{black}An example of the initial placement strategy for the circuit in Fig.~\ref{fig:Dependency Graph}}}
    \label{fig:DFS placement}
\end{figure}

{\color{black}This strategy attempts to first sort the logical qubits in a depth-first-search (DFS) order and then map them to the physical qubits according to the DFS order on the Device Graph. Algorithm \ref{alg: dfs} shows the procedure of sorting the logical qubits in the DFS order on the quantum circuit. Let’s use the circuit on Fig.~\ref{fig:Dependency Graph-a} again as an example. Assume the list of the double-qubit gates (i.e. DQG) are initialized to { $G1, G2, G3, G4, G5, G6, G7$ }. We will first pick $G1$ (Line 3) and one of its inputs $q_3$ (Line 4) to call the \textbf{DFS\_RECUR} subroutine (Line 5). In the subroutine, $q_3$ will be added to DFS (Lines 11-14) and, $q_4$, the first qubit on the adjacency list of $q_3$, will be passed to the recursive call (Line 17). Continuing with this pre-order traversal procedure, we can derive the DFS order of the logical qubits as $\{$$q_3, q_4, q_1, q_0, q_6, q_5, q_2$$\}$. On the other hand, the DFS order of the physical qubits in Fig.~\ref{fig:ring layout} can be easily computed as $\{$$Q_0,Q_1,Q_2,Q_3,Q_4,Q_5,Q_6,Q_7$$\}$. Therefore, the initial qubit mapping will be $\{$($q_3, Q_0$), ($q_4, Q_1$), ($q_1, Q_2$), ($q_0,Q_3$), ($q_6, Q_4$), ($q_5, Q_5$), ($q_2, Q_6$), ( , $Q_7$)$\}$, as shown in Fig.~\ref{fig:placement result}. 

Note that with this approach, the inputs of the first three double-qubit gates $G1, G2$, $G3$, and even the fourth double-qubit gate $G4$, will be placed in adjacency (i.e. ($Q_0, Q_1$), ($Q_5, Q_4$), ($Q_3, Q_2$), and ($Q_2, Q_1$)). Therefore, no SWAP insertion is needed. However, for the next double-qubit gate $G5$, although its inputs seem next to each other in the circuit (i.e. $q_1$ and $q_2$), they are apart in terms of the physical qubits (i.e. $Q_2$ and $Q_6$) and thus a sequence of SWAP insertions will be performed. 

Clearly, the closeness of the inputs of the subsequent double-qubit gates (i.e. $G6$ and $G7$) will be affected by the SWAP insertions of $G5$. Therefore, it is impossible for the initial placement strategy to ensure the overall optimality of the qubit mapping problem. Our observation indicates that initial placement tends to play a more pivotal role in shallow circuits, where the qubits interact only with a limited number of others. However, as the circuit complexity grows, the influence of the initial placement is confined to the first few ``layers'' before swapping paths disturbing the initial placement, and thus the routing and scheduling strategies begin to exert a more dominant influence on the final outcome. We examine our speculation in Section \ref{section:initialPlacementExp}.}

\begin{algorithm}
\caption{DFS$\_$TRAVERSAL}\label{alg:dfs}
\begin{algorithmic}[1]

\Procedure{DFS$\_$TRAVERSAL}{$DQG$}
\Comment{$DQG$ denote the list of double-qubit gates}
\State $DFS = \emptyset$
\For{gate $\in$ $DQG$}
\For{$q$ $\in$ gate}  
\Comment{$q$ is the input qubit}
\State DFS$\_$RECUR($q$, $DFS$)
\EndFor
\EndFor
\State \textbf{return} $DFS$
\EndProcedure

\Procedure{DFS$\_$RECUR}{$q,DFS$}
\If{$q$ is visited} \State \textbf{return}\EndIf
\State $q\rightarrow $ setVisited() 
\State $DFS\leftarrow q$
\For{$q' \in q.adj\_list$}
% \Comment{$q.adj\_list$ is the adjacency list of $q$}
\State DFS$\_$RECUR($q',DFS$)
\EndFor
\EndProcedure
\end{algorithmic}
\label{alg: dfs}
\end{algorithm}

\subsection{Routing: Duostra (Dual-source Dijkstra)}
In this paper, we present a novel routing algorithm called Duostra (Dual-source Dijkstra) for handling a double-qubit gate that operates on physically distant qubits. The algorithm outputs an optimal routing path and provides a SWAP sequence that efficiently brings the two involved qubits, of a double-qubit gate, adjacent to each other. Additionally, we demonstrate that the time complexity of this algorithm is $O(n\log n)$, where $n$ represents the number of qubits in the quantum device.

In the following, we first introduce two important terminologies, ``occupied time'' and ``routing path'', for our routing algorithm. {\color{black}We then describe the main novelty of the routing procedure} and present the details of the algorithm in the end.

\subsubsection{Terminologies}\label{section:terminologies}
\begin{definition}[Occupied Time]
We have adopted the concept introduced in \cite{deng_codar_2020} to define the notion of ``occupied time” for a qubit. When a physical qubit is involved in an operation, such as SWAP or CX, it is considered occupied. If an operation is carried out on a physical qubit $Q_n$ from time $\tau_0$ to $\tau_1$, we define the occupied time of $Q_n$ as the finishing time of that operation, denoted as $Q_n.ocp = \tau_1$. This means that the qubit cannot be assigned to another operation until time $\tau_1$.
\end{definition}

We further extend the concept of occupied time to the edge of the Device Graph. Let $e(i, j)$ be an edge that connects two adjacent qubits $Q_i$ and $Q_j$ on $DevG$. Applying a SWAP gate on this edge will swap the logical qubits that $Q_i$ and $Q_j$ applied on. We then calculate the occupied time for the edge with or without applying SWAP by:
\begin{align} 
     e(i, j)_{swap}.ocp &\triangleq \max (Q_i.ocp, Q_j.ocp) + \tau_{swap}  \  &\text{with SWAP} \label{Eq: with_SWAP} \\
     e(i, j).ocp &\triangleq \max (Q_i.ocp, Q_j.ocp) \  &\text{without SWAP} \label{Eq: without_SWAP}
\end{align}    
where $Q_i.ocp$, $Q_j.ocp$ and $\tau_{swap}$ represent the occupied time of $Q_i$, occupied time of $Q_j$, and the operating time for a SWAP gate (i.e. 6 units as defined in \cite{deng_codar_2020}).

\begin{definition}[Routing Path]
\label{def: routing path}
In the context of a source qubit $Q_s$ and a distant target qubit $Q_t$, we represent the routing path $RP(s, t)$ as a pathway from $Q_s$ to $Q_t$, which involves a sequence of SWAP operations applied along the edges of this path.
\end{definition}

Let $Q_1$, $Q_2$,... , $Q_n$ be the $n$ qubits between $Q_s$ and $Q_t$ along the routing path $RP(s, t)$. In other words, $Q_s$ and $Q_1$, $Q_i$ and $Q_{i+1}$ (for $i$ between $1$ and $n-1$), and $Q_n$ and $Q_t$ are all adjacent. With Eq.~\eqref{Eq: with_SWAP}, we can recursively calculate the occupied time of a routing path as

\begin{equation*} %\label{Eq:routingPath}
    \begin{split}
    RP(s, t).ocp &= \max(RP(s, n), e(n, t)_{swap}.ocp) \\
        &= \max(\max(RP(s, n\!-\!1), e(n\!-\!1, n)_{swap}.ocp),e(n, t)_{swap}.ocp) = ...,
    \end{split}
\end{equation*} 
where $e(n-1, n)$ and $e(n, t)$ are adjacent edges between $Q_{n-1}$ and $Q_n$, and $Q_{n}$ and $Q_t$, respectively.

During the computation of the occupied time for a routing path, the occupied time of the qubits along that path is updated accordingly. 

\begin{figure}
     \centering
     \begin{subfigure}[b]{0.49\textwidth}
         \centering
         \includegraphics[width=\textwidth]{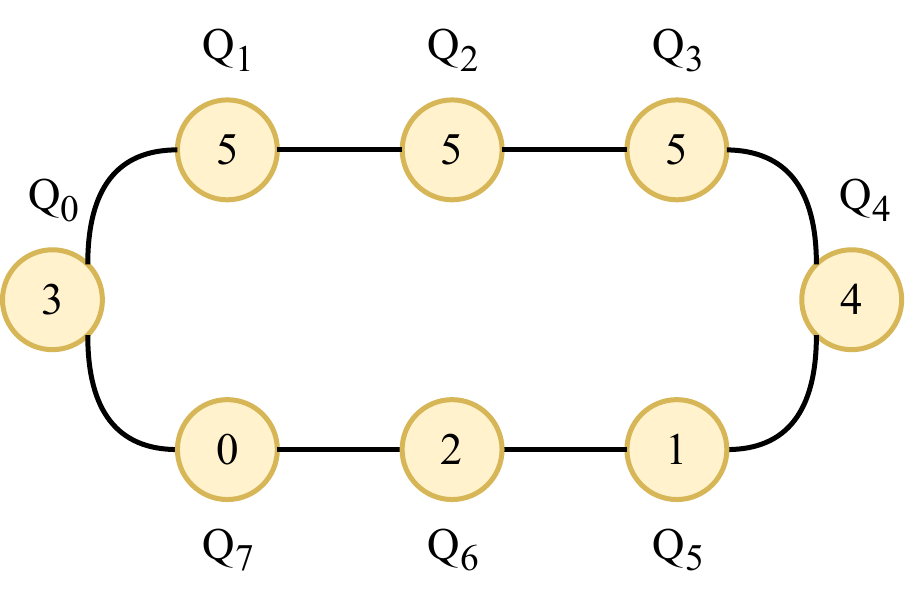}
        \caption{{\color{black}Occupied time of the qubits}}
         \label{fig:Duostra-a}
     \end{subfigure}
     \hfill
     \begin{subfigure}[b]{0.49\textwidth}
         \centering
         \includegraphics[width=\textwidth]{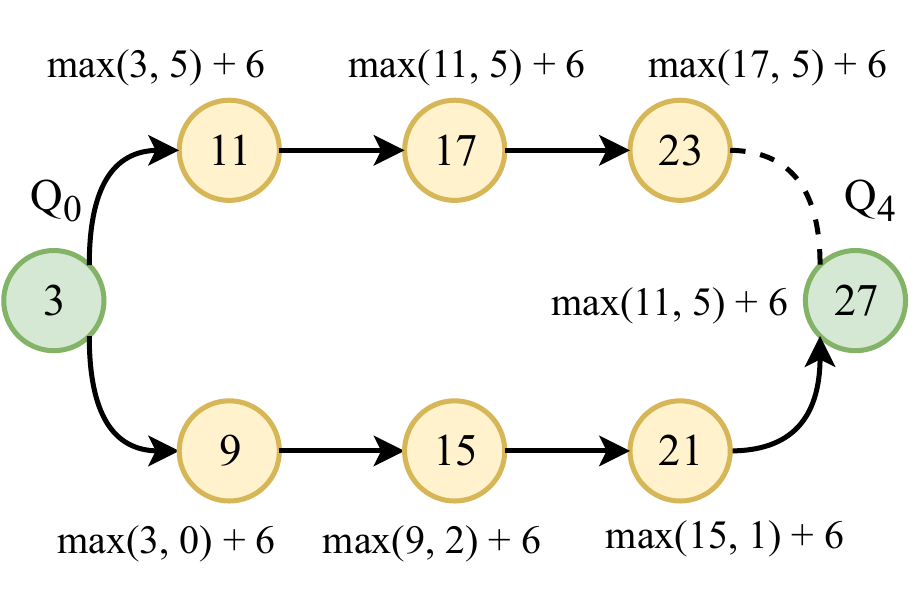}
        \caption{{\color{black}Occupied time of the routing path RP(0,4)}}
        \label{fig:Duostra-b}
     \end{subfigure}
     
     \begin{subfigure}[b]{0.49\textwidth}
         \centering
         % \vspace{10px}
         \includegraphics[width=\textwidth]{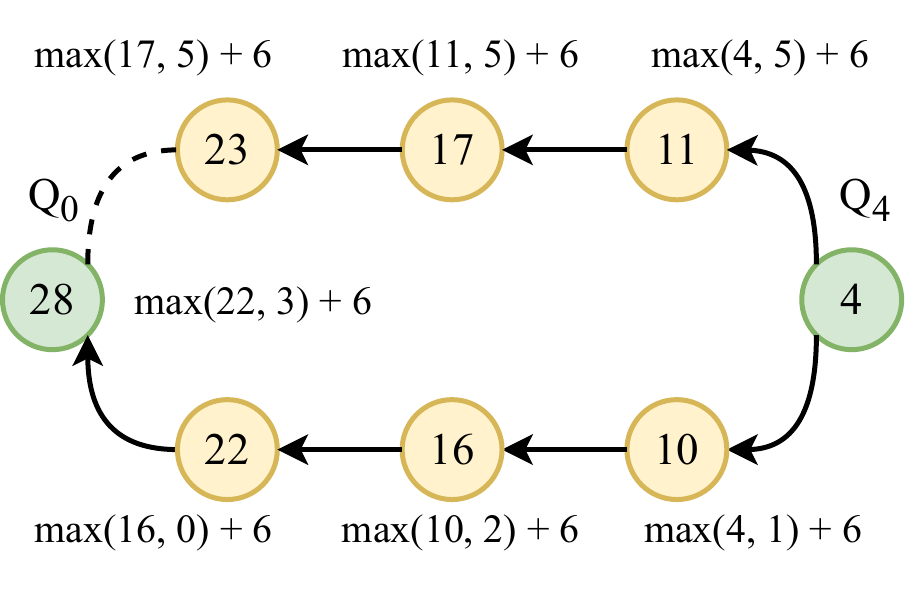}
        \caption{\color{black}Occupied time of the routing path RP(4, 0)} 
         \label{fig:Duostra-c}
     \end{subfigure}
     \hfill
     \begin{subfigure}[b]{0.49\textwidth}
         \centering
         \includegraphics[width=\textwidth]{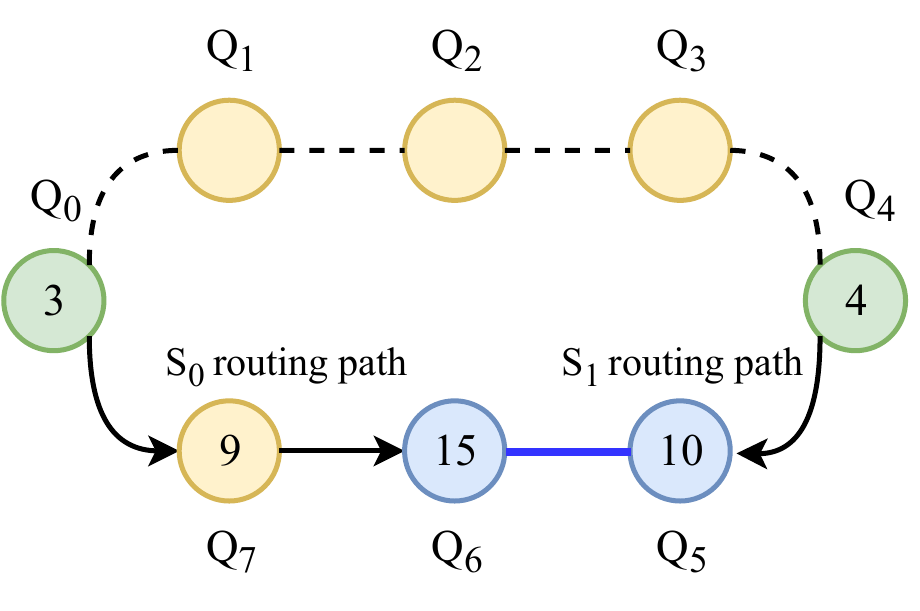}
        \caption{{\color{black}Optimal routing paths RP(0,6) and RP(4,5)}}
        \label{fig:Duostra-d}
     \end{subfigure}
     \caption{Demonstration of the Duostra Procedure {\color{black} for the routing of the double-qubit gate $G6$}}
     \label{fig:Duostra}
\end{figure}

{\color{black}To illustrate this concept, we proceed with the example outlined in Section \ref{section:initialPlacement}. After initial mapping, we can execute the single-qubit gates (i.e. H, Z, S, X, T) and the two-qubit gates $G1, G2, G3$, and $G4$ without the need of SWAP insertions. The occupied time for the physical qubits after the executions is illustrated in Fig.~\ref{fig:Duostra-a} and according to the dependency graph in Fig.~\ref{fig:Dependency Graph-b}, we have the double-qubits gates $G5$ and $G6$ on the Waitlist. Let’s assume we pick $G6$ for the next execution. Since its inputs are now placed on two non-adjacent qubits $Q_0$ and $Q_4$, we need to find a routing path to bring them to adjacency.}

{\color{black}
If we choose $Q_0$ as the source and $Q_4$ as the target, we observe that the SWAPs can be performed along two distinct paths: $(Q_0, Q_1, Q_2, Q_3, Q_4)$ or $(Q_0, Q_7, Q_6, Q_5, Q_4)$. } {\color{black}The occupied time for both of these routing paths and their corresponding qubits is calculated as in Fig.~\ref{fig:Duostra-b}. Upon analysis, we find that the occupied time of the path $(Q_0, Q_1, Q_2, Q_3, Q_4)$ (i.e., $29$) is greater than that of the path $(Q_0, Q_7, Q_6, Q_5, Q_4)$ (i.e., $27$). As a result, in our efforts to optimize the execution time of the quantum circuit, we can disregard the former path for future considerations.} In general, when dealing with a source qubit $Q_s$ and a target qubit $Q_t$, their optimal routing path is denoted as $RP_{opt}(s, t)$. For this specific example, {\color{black} $RP_{opt}(0, 4) = 27$.

It is important to note that if we interchange the roles of the source and target qubits for $Q_0$ and $Q_4$, the occupied time for the routing path and qubits will differ. In Fig.~\ref{fig:Duostra-c}, we present the calculation of the occupied time for the routing paths and qubits when taking $Q_4$ as the source and $Q_0$ as the target. Notably, $RP_{opt}(4, 0)$ is found to be $28$, which is larger than $RP_{opt}(0, 4)$ (i.e. $27$). Therefore, choosing $Q_0$ as the source and $Q_3$ as the target proves to be the more favorable routing path.}

\begin{figure}
     \centering
     \begin{subfigure}[b]{0.49\textwidth}
         \centering
         \includegraphics[width=\textwidth]{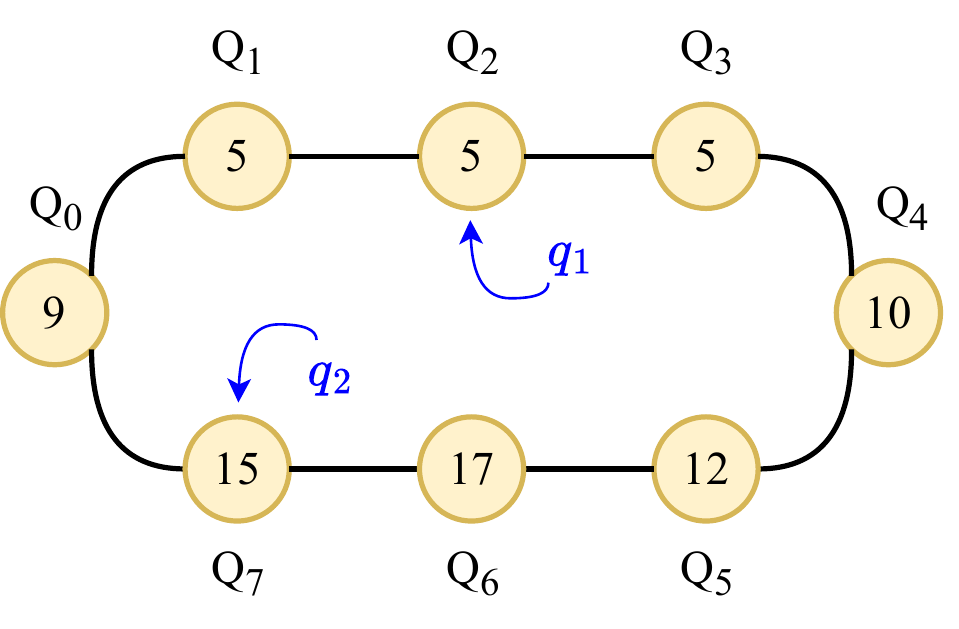}
        \caption{{\color{black}Occupied time of the qubits after applying $G6$}}
         \label{fig:Duostra-next-a}
     \end{subfigure}
     \hfill
     \begin{subfigure}[b]{0.49\textwidth}
         \centering
         \includegraphics[width=\textwidth]{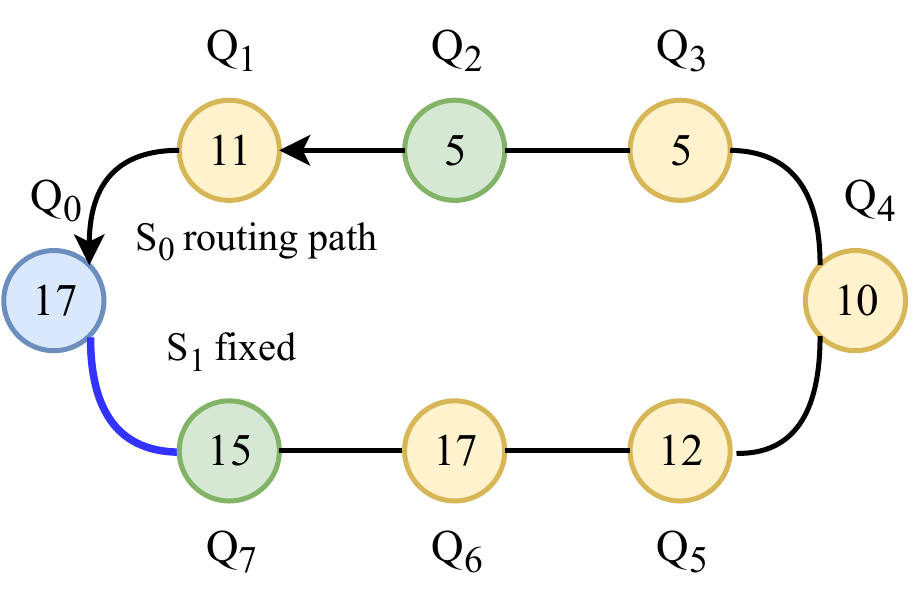}
        \caption{{\color{black}Optimal routing path RP(2,0)}}
        \label{fig:Duostra-next-b}
     \end{subfigure}
    \caption{{\color{black}Demonstration of the Duostra Procedure for the routing of the double-qubit gate $G5$}}
     \label{fig:Duostra-next}
\end{figure}

{\color{black}Furthermore, if we treat both $Q_0$ and $Q_4$ as sources and search for two routing paths ($Q_0$ to $Q_4$ and $Q_4$ to $Q_0$) between them, we can observe that when both routing paths converge at the edge $e(5, 6)$, an optimal routing path with a minimal occupied time of $15$ is achieved (Fig.~\ref{fig:Duostra-d}). In simpler terms, by swapping the logical qubits on $Q_0$ to $Q_6$ and on $Q_4$ to $Q_5$, we can bring the computation of the double-qubit gate to a pair of adjacent qubits and attain the most optimal qubit mapping for it. This leads us to }the dual-source routing algorithm, named ``Duostra,'' as depicted below:

Given a double-qubit gate that operates on two physically remote qubits (i.e. the sources) $Q_{s_0}$ and $Q_{s_1}$, the Duostra algorithm aims to find a pair of adjacent qubits (i.e. the targets) $Q_{t_0}$ and $Q_{t_1}$ on the Device Graph in such a way that the maximum of $RP_{opt}(s_0, t_0).ocp$ and $RP_{opt}(s_1, t_1).ocp$ can be minimized. That is:
\begin{equation}\label{eq:target}
    \min_{e \in E}(\max (RP_{opt}(s_0, t_0).ocp, RP_{opt}(s_1, t_1).ocp)).
\end{equation}

\subsubsection{The novelty of Duostra algorithm}

\begin{figure}[h]
    \begin{subfigure}[b]{0.45\textwidth}
    \centering
    \includegraphics[width=0.9\textwidth]{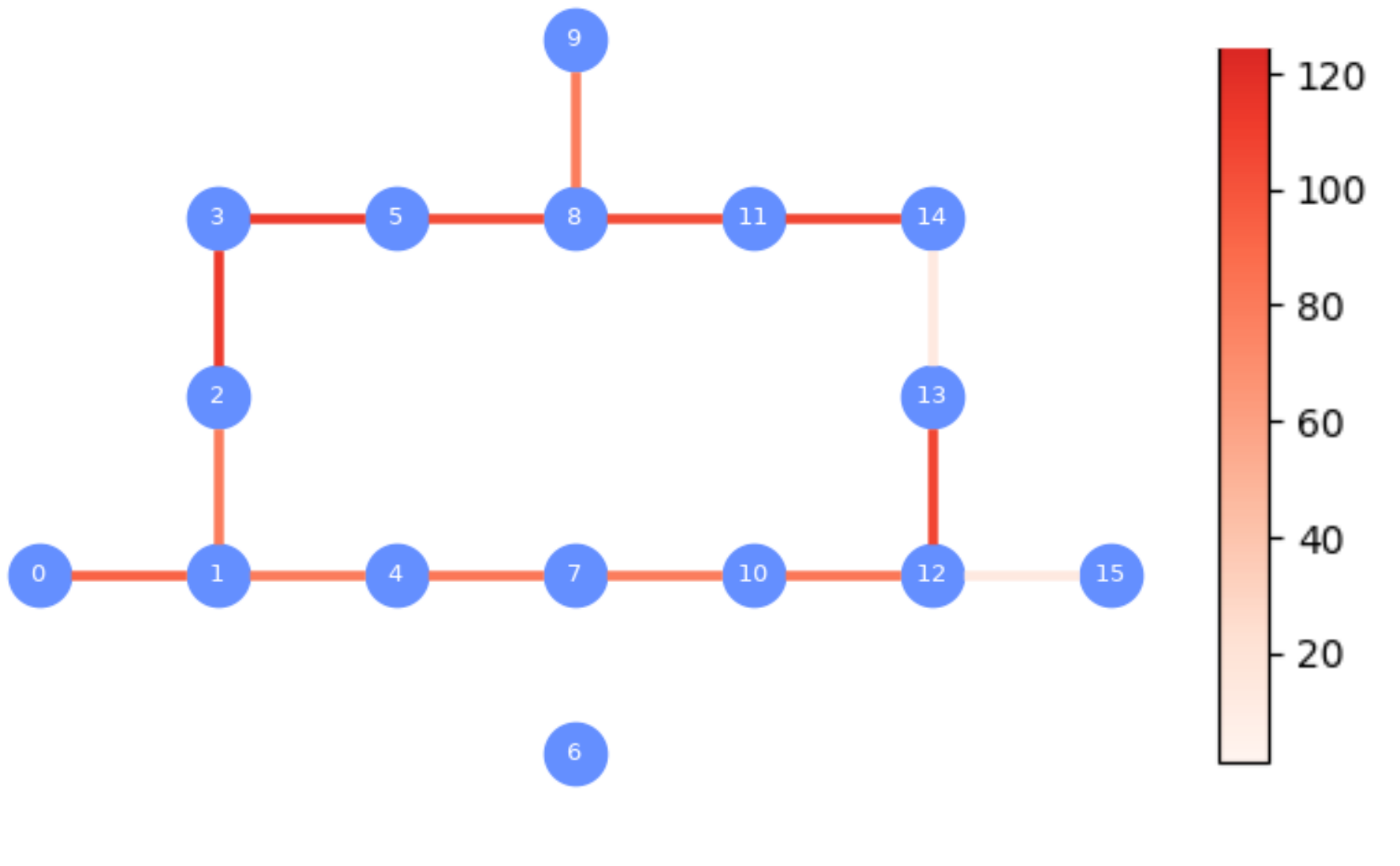}
    \caption{{\color{black}Edge utilization of Duostra router with standard deviation 35.26}}
    \label{fig:duostra heatmap}
    \end{subfigure}
    \hfill
    \begin{subfigure}[b]{0.45\textwidth}
    \centering
    \includegraphics[width=0.9\textwidth]{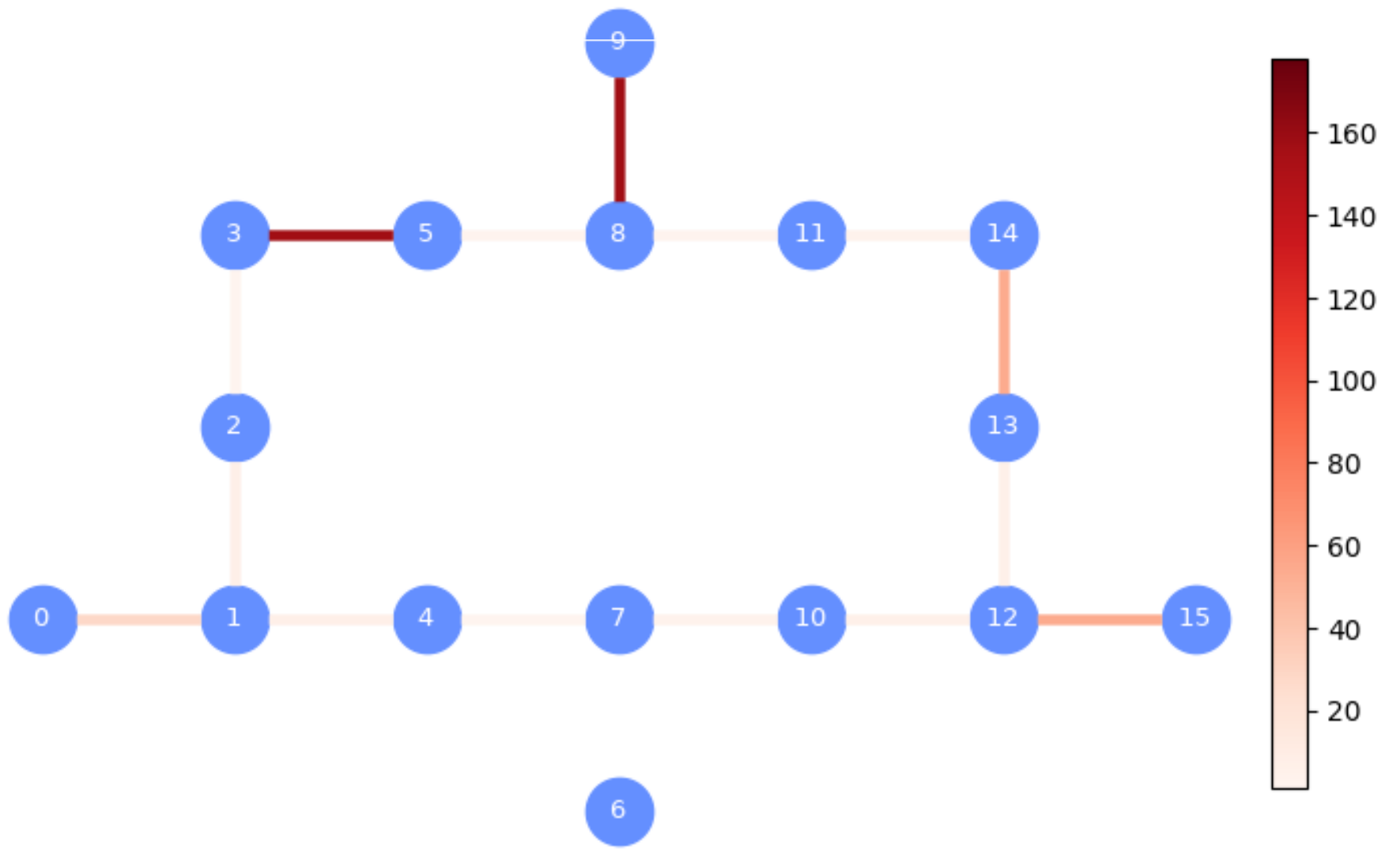}
    \caption{{\color{black}Edge utilization of Shortest path router with standard deviation 67.00}}
    \label{fig:Shortest heatmap}
    \end{subfigure}
    \caption{{\color{black}Heat graphs of SWAP counts for each edge between different routers on benchmark \textit{cm42a$\_$207} and device \textit{ibmq\_guadalupe}. Both of them are normalized to the same gradients.}}
    \label{fig:heat}
\end{figure}

{\color{black} The Duostra algorithm is a modified version of the classic shortest-path finding algorithm, i.e. the Dijkstra’s algorithm \cite{dijkstra_note_1959}, with the key differences in traversing the search tree from dual sources and dynamically updating the availabilities of the nodes (i.e. the occupied time of the physical qubits) during the search. Let’s continue the qubit mapping for the previous example. After we apply SWAPs between ($Q_0, Q_7$), ($Q_7, Q_6$) and ($Q_5, Q_4$), the double-qubit gate $G6$ can now be executed on the adjacent physical qubits ($Q_6$, $Q_5$). At the same time, the inputs of the other double-qubit gate $G5$ are thereby swapped to $Q_7$ and $Q_2$, and the occupied time of the physical qubits are then updated as depicted in Fig.~\ref{fig:Duostra-next-a}\footnote{{\color{black}We increment the occupied time for $Q_6$ and $Q_5$ by 2 due to the execution of the CX gate.}}. 

Note that if we apply the traditional shortest path approach to resolve the coupling constraint of $G5$, it is intuitive that we will execute a SWAP between $Q_0$ and $Q_7$, and another SWAP between $Q_1$ and $Q_2$. This will lead to the occupied time 21 as a result.

However, if we perform the Duostra algorithm, as shown in Fig. \ref{fig:Duostra-next-b}, by considering and dynamically updating the occupied time of the qubits on the routing path, we will keep one of the sources unmoved on $Q_7$, and swap the other source from $Q_2$ to $Q_0$. This will generate the optimal mapping with the occupied time 17.

To further elucidate the differences above, let’s compare the number of SWAPs executed on the edges by the Duostra and the traditional Shortest-Path (SP) algorithms as in Fig.~\ref{fig:heat}. The result on the benchmark circuit $cm42a\_207$ shows that the SWAP operations by the SP algorithm are not applied uniformly. It has the standard deviation 67.00 and the maximum number 178. On the other hand, Duostra renders a much balanced distribution of the SWAP operations. Its standard deviation is 35.26 and the maximum number is 112. This discrepancy suggests that Duostra tends to choose the routing path with smaller occupied time and thus yield more balanced and parallel results. It can thereby reduce the mapping cost.}

\subsubsection{Detailed description of the algorithm}
We present the pseudo-code of the Duostra algorithm in Algorithm \ref{alg:duostra} below. To begin, we initialize a priority queue, $PQ$, and mark all vertices $v\in V$ as unseen and unvisited. The input sources of the routing problem, $s_0$ and $s_1$, are then marked as seen and visited. For each vertex $v$, we utilize a field $v.source$ to indicate its routing source. As there are only two possible routing sources, $s_0$ and $s_1$, we set the routing sources of $s_0$ and $s_1$ as themselves (Lines 2 and 3). Subsequently, we add all the adjacent vertices of $s_0$ and $s_1$ to $PQ$, updating their occupied time, marking them as seen, and recording their sources as either $s_0$ or $s_1$ (Lines 5 and 6).

\begin{algorithm}
\caption{Duostra (Dual-source Dijkstra)}\label{alg:duostra}
\begin{algorithmic}[1]

\Procedure{Duostra}{$Graph, s_0, s_1$}
\State $s_0.source$ $\gets$ $s_0$
\State $s_1.source$ $\gets$ $s_1$
\State $PQ\gets$ priority queue
\State pushUnseenNeighbors$(PQ,s_0)$
\State pushUnseenNeighbors$(PQ,s_1)$
\While{$PQ$ is not empty}
\State $m\gets$ nodeWithLowestCost($PQ$) \Comment{Cost is the occupied time of the vertex}
\State $m.visited \gets True$
\If{$\exists$ visited neighbor $v$ s.t. $m.source \not =v.source$}
\State $RP(m.source, m)\gets $ backtrace$(m)$
\State $RP(v.source, v)\gets $ backtrace$(v)$
\State \textbf{return} $(path_{s_0}, path_{s_1})$
\EndIf 
\State pushUnseenNeighbors$(PQ,m)$
\EndWhile\label{euclidendwhile}
\EndProcedure

\Procedure{pushUnseenNeighbors}{$PQ,m$}
\For{$v\in$ Neighbor$(m)$ and $v.unseen$}
\State $v.source \gets m.source$
\State $v.seen \gets True$
\State $PQ$.push$(v)$
%\Comment{Insert order:$\min(Cost(s,v),Cost(t,v))$}
\EndFor
\EndProcedure
\end{algorithmic}
\end{algorithm}

Next, we proceed iteratively by selecting the vertex with the smallest occupied time, $m$, from $PQ$ and marking it as visited (Lines 8 and 9). If there exists an adjacent vertex $v$ of $m$ that has been visited and is routed from the other source different from $m.source$ (Line 10, i.e. $v.source \neq m.source$), then we conclude the procedure and establish an optimal routing solution. This solution includes the optimal routing path from $m.source$ to $m$ (i.e. $RP_{opt}(m.source, m)$), the optimal routing path from $v.source$ to $v$ (i.e. $RP_{opt}(v.source, v)$), and the edge $e(m, v)$ (see Section \ref{section:ProveDuostra} for the proof of the optimality of the Duostra algorithm). Otherwise, we add all the unseen neighbors of $m$ to $PQ$ and continue with the iteration. {\color{black}Since the Duostra algorithm can be viewed as a Dual-source Dijkstra algorithm with dynamic occupied time as the cost function, the algorithm has the time complexity $O(n\log n)$, where $n$ represents the number of physical qubits.}

It is noteworthy that there is no need to maintain separate priority queues for the search trees originating from the dual sources $s_0$ and $s_1$. Instead, all seen vertices can be recorded in a single priority queue, irrespective of whether their sources are $s_0$ or $s_1$. As we define the cost of a seen vertex $v$ in the priority queue as the occupied time of the optimal routing path from $v.source$ to $v$, the top vertex in the priority queue, with the lowest cost, represents the earliest qubit that can be operated using a SWAP gate.

Furthermore, once a vertex is pushed into the priority queue, its cost remains unchanged. As a result, we can ensure that when the two search trees from $s_0$ and $s_1$ converge at a pair of adjacent physical qubits, they will indeed form the optimal routing path with the smallest occupied time. This is because if there were an alternative sequence that could be completed earlier, it would have been visited from the priority queue and halted the search process earlier. The proof of the optimality of the priority queue is provided in Section \ref{section:ProveQueue}.

\subsection{Scheduling: Limitedly-Exhaustive (LE) Search and Shortest-Path (SP) Estimation}
Given a quantum circuit, the scheduler plays a crucial role in determining the optimal execution sequence of the gates on the quantum computing device, with the objective of optimizing the total execution time or achieving other specific goals. It is important to note that, at any given moment, we have the flexibility to select any gate from the Waitlist for the subsequent execution and then employ the Duostra algorithm to find the most optimal routing path.

Once a gate is executed and SWAPs are performed, the occupied times of the relevant qubits are updated, and the subsequent gates on the Dependency Graph are added to the Waitlist. This procedure refers to the scheduler routine, as shown in Fig.~\ref{fig:framework}. Since different execution orders can lead to varying total execution time, to attain the global optimum of the scheduling problem, we must consider all possible execution sequences of the gates (i.e., different gate selections from the Waitlist). However, this exhaustive enumeration proves to be an intractable task \cite{siraichi_qubit_2018}, even for relatively small-scale quantum circuits.

In this paper, we propose two heuristic approaches that adopt different subsets of the execution sequences at each step when choosing the gate to route for subsequent execution. These heuristic methods provide practical and effective strategies for addressing the scheduling problem in the context of quantum circuits.

\subsubsection{Limitedly-Exhaustive (LE) Search}
As exhaustively enumerating all possible gate sequences on the Dependency Graph proves to be exponentially costly, we introduce a Limitedly-Exhaustive Search approach to constrain the search depth for each decision. In this method, the user specifies a depth limit $d$, and the search process focuses on exhaustively exploring all depth-$d$ sequences originating from any gate in the Waitlist. For each of these depth-$d$ sequences, the Duostra routing algorithm is utilized to calculate their corresponding occupied time.

The scheduler then proceeds to select the sequence with the minimal occupied time, determining its corresponding gate in the Waitlist for execution. Subsequently, the Waitlist is updated, and the process is repeated for the subsequent depth-$d$ sequences. To illustrate, if the depth limit is set to $2$, the Limitedly-Exhaustive Search will first enumerate all the gates in the Waitlist, finding the optimal routing path for each gate using Duostra, and updating the corresponding occupied time and Waitlists accordingly. Next, for each gate in the previous iteration and its updated Waitlist, the gates in the updated Waitlist are enumerated, and the above process is repeated again. Finally, the gate that results in the smallest occupied time among these depth-$2$ sequences will be selected for execution, and the Limitedly-Exhaustive search process will be restarted. This approach offers an efficient strategy to tackle the scheduling problem by limiting the search depth while reducing computational complexity.

\subsubsection{Shortest-Path (SP) Estimation} \label{section:SPEst}
While the Limitedly-Exhaustive (LE) Search heuristic may provide an approximation close to the global optimum solution when employing a large search depth $d$, its practical applicability remains limited due to the extensive calculations involved in the path enumeration. Consequently, it can only be effectively applied to small-scale designs. To address this challenge and being inspired by Li \textit{et al}. \cite{li_tackling_2019}, we propose an alternative heuristic approach that considers the concept of the shortest path \cite{floyd_algorithm_1962} between the physical qubits of the gate.

The SP-Estimation goes through all the gates in the Waitlist and selects the one with the minimal value of this equation:
\begin{equation}\label{eq:scheduling}
    \begin{aligned}
         \max(g_i.Q_i.ocp, g_i.Q_j.ocp) + c\cdot SP (g_i.Q_i, g_i.Q_j),
    \end{aligned}
\end{equation}
where $g_i$ is a double-qubit gate from the Waitlist, $g_i.Q_i$ and $g_i.Q_j$ denote its current qubit assignments, $SP(g_i.Q_i,g_i.Q_j)$ indicates the shortest path between these qubits, and $c$ is a constant factor to {\color{black} take} the execution time of the shortest path {\color{black}into consideration}. In our implementation, we set $c$ to be $1$. After selecting a gate with a minimal value of Eq. \eqref{eq:scheduling} from the Waitlist, we employ the Duostra algorithm to calculate the occupied time of the circuit, and then continue with the updated Waitlist. 

It is essential to emphasize a significant difference between the Duostra routing path and the shortest-path routing path. If we were to directly choose the shortest path as the routing path, collisions may occur where the route conflicts with the swaps of another gate. This situation would force us to pause the swapping procedure, leading to a prolongation of the execution time. On the contrary, the Duostra algorithm takes into account the occupied time of each qubit, allowing it to avoid busy routes and efficiently achieve the optimal path for the gate. This distinction is critical in optimizing the overall Mapping Cost (c.f. Section \ref{def:cost}) of the quantum circuit, see {\color{black}experiments in} Section \ref{section:Futurism}. Therefore, given the shortest path, we still need to rely on the Duostra router to produce the local optimal result.

The time complexities of the LE Search approaches and the SP Estimation are $O(N^2n)$ and $O(N(W + R))$, respectively. Here, $N$ represents the number of gates in the circuits, $W$ is the number of gates in the Waitlist, $R$ stands for the time complexity of the router, and $n$ indicates the number of qubits.

In summary, the LE Search approach explores all possible gate sequences up to a specified depth $d$. It has the capability to close in on global optimum solutions for smaller designs when the depth $d$ is set sufficiently large. Conversely, the SP Estimation approach prioritizes better scalability, even if it may sacrifice some performance. As a result, the SP Estimation approach can be effectively applied to larger circuits, thus paving the way toward achieving quantum advantage.

\section{Proof of Duostra Optimality  }\label{section:ProveDuostra}
We design the Duostra Algorithm by capitalizing on the property of local optimality in order to strike a balance between the mapping cost and the program runtime for large circuits. Given a double-qubit gate whose control bit is not adjacent to its target bit, Duostra Algorithm can generate a SWAP sequence with minimum Mapping Cost. We will prove the Duostra Optimality as follows. 

In our model, physical qubits are represented as vertices, and given a vertex $v_i$, the parent $v_{p_i}$ of $v_i$ is the vertex adjacent to $v_i$ with the smallest occupied time. In other words, $v_{p_i}$ is the previous vertex of $v_{i}$ in the routing path. Thus, according to Eq. \eqref{Eq: with_SWAP}, the occupied time of $v_i$ under the condition of the swap edge $e(v_{i}, v_{p_i})_{swap}$ can be calculated as:

\begin{equation}\label{pf01}
    \begin{aligned}
        v_i.ocp\big\rvert_{e(v_{i}, v_{p_i})_{swap}} = \max(v_{i}.ocp, v_{p_i}.ocp) + \tau_{swap}.
    \end{aligned}
\end{equation}

In our Duostra Algorithm as presented in Algorithm \ref{alg:duostra}, the vertices are categorized into three types: visited ($t$), seen ($s$), and unseen vertices ($u$). Both seen and unseen vertices are all unvisited. We call the control and the target bits of the double-qubit gate the routing sources. At first, we mark the routing sources as visited, and their adjacent vertices as seen. During each iteration of the algorithm, we select a vertex from the set of seen vertices and mark it as visited. Following this, the unseen vertices connected to the chosen visited vertex are then labeled as seen.

\begin{lemma}\label{section:ProveQueue}
    The seen set contains the unvisited vertex with the minimum occupied time
\end{lemma}

We prove lemma \ref{section:ProveQueue} as follows. Since an unseen vertex $u_j$ must be connected to the seen set $S$ by at least one edge, let the seen vertex connect through this edge be $s_{arb}$, and we can calculate $u_j.ocp$ after applying the SWAP gate by Eq. \eqref{pf01} as
\begin{equation}\label{pf02}
    \begin{aligned}
         u_j.ocp\big\rvert_{e(u_{j}, s_{arb})_{swap}} = \max(u_{j}.ocp, s_{arb}.ocp) + \tau_{swap} 
         % > s_{arb}.ocp \geq \min_{s_i\in S}(s_i.ocp).
    \end{aligned}
\end{equation}
Based on Eq.~\eqref{pf02}, we observe that as long as there is one edge between $u_{j}$ and the seen set, it would lead to $u_{j}.ocp\rvert_{e(u_{j}, s_{arb})_{swap}} > s_{arb}.ocp$. Therefore, since the occupied time of any seen vertex ($s_{arb}.ocp$) is greater than or equal to the minimum occupied time of all the seen vertices, the occupied time of an unseen vertex will inevitably be greater than the vertex with the smallest occupied time in the seen set.

We can turn to prove the Duostra Optimality with lemma~\ref{section:ProveQueue}. In the routing problem, our objective is to find the best edge $e^*$ between the best adjacent pair, $v^*_0$ and $v^*_1$, which minimizes the objective function, defined in Eq. \eqref{eq:target}, $$e(v^*_0, v^*_1)^*_{swap} = \max(RP(r_0, v^*_0).ocp, RP(r_1, v^*_1).ocp),$$ where $r_0$ and $r_1$ are the given routing sources. To minimize the objective function, we compare the occupied time of all the seen vertices generated by the two routing paths and select the vertex with the minimum occupied time. This vertex will prolong one of the routing paths. Therefore, by iteratively searching for the optimal solution, we come to the Duostra Optimality.

\begin{lemma}[Duostra Optimality]\label{lemma:Optamility}
    The first edge where the two routing paths converge is the solution~$e^*$, which has the smallest occupied time.
\end{lemma}

We prove the lemma \ref{lemma:Optamility} by contradiction. Suppose Duostra Algorithm identifies a sub-optimal edge $e^-$ with two vertices, $v^-_0$ and $v^-_1$, then the occupied time of $e(v^-_0, v^-_1)_{swap}$ will be:
\begin{equation}
\begin{aligned}
     e(v^-_0, v^-_1)_{swap} = \max(RP(s_0, v^-_0).ocp, RP(s_1, v^-_1).ocp) > \max(RP(s_0, v^*_0).ocp, RP(s_1, v^*_1).ocp).
\end{aligned}
\end{equation}
Because we only prolong the routing paths by selecting a vertex with the minimum occupied time, the sub-optimal solution $e^-$ with the bigger occupied time must be found after the optimal solution $e^*$. As a result, Duostra Algorithm finds the optimal solution given a double-qubit gate, thus ensuring efficient and effective qubit mapping in quantum circuits.

\section{Experimental Results} \label{section:EXPERIMENT}
In this section, we first present our experiment setup and an experiment to testify to the significance of initial placement. Following that, we respectively demonstrate experiments on mid-size circuits to showcase our robustness on both Mapping Cost and runtime, on large circuits to witness the scalabilities, and on extra-large circuits to prove our readiness for future quantum {\color{black}advantage}. We conducted all experiments on a 13th Gen Intel\textregistered\  Core\texttrademark\  i9-13900K with a total of 126G memory.
% Copyright: \copyright Math-Linux.Com\\
% Trademark: \texttrademark or \textsuperscript{TM} Math-Linux.Com  \\
% Registered: \textregistered Math-Linux.Com \\
\subsection{Experiment Setup}

\subsubsection{Compared Methods}\label{section:comparedMethod}
{\color{black}We adopt four existing tools for the comparison in our experiments, namely QMAP \cite{wille_mqt_2023}, TOQM \cite{zhang_time-optimal_2021}, t$|$ket$\rangle$ \cite{sivarajah_tket_2021}, and IBM Qiskit \cite{aleksandrowicz_qiskit_2019} (version 1.0.1) with default and SABRE mappers.} For QMAP and TOQM, we use their heuristic versions because the exact versions are impractical for our experiments, as it takes them over a day to generate the mapping results. On the other hand, to ensure the fair comparison of the qubit mapping results, we set the optimization level of Qiskit to 0 so that it will not perform additional circuit optimization. Other than these four tools, we also implement a baseline version of our framework by replacing the Duostra router with a shortest-path router, which directly takes the shortest path as the swap path.

\subsubsection{Addressing the connectivity challenges across the NISQ era and beyond}

{\color{black}
Our proposed qubit mapping framework is not limited to specific devices in NISQ era. However, based on current knowledge, maintaining a low average node degree in superconducting quantum devices, typically less than three, is crucial to minimize frequency collisions and crosstalk errors \cite{chamberland_topological_2020}. For example, IBM's 433-qubit Osprey proposed in 2022 and the 1121-qubit Condor unveiled in 2023 adhere to the heavy-hexagon topology. Therefore, we opt for the heavy-hexagon structure in our experiment setup to align with the current understanding of the connectivity issues, although our research is not confined to the hexagon topology.

Looking ahead to the era beyond NISQ, if the qubit fidelity improves sufficiently, we can directly address the qubit mapping problem when executing the quantum circuits on the quantum devices. Therefore, our algorithm can continue to prevail. However, if the quantum error correction codes (QECC) are needed, we can view the quantum error correction as an amalgamation of the correction code circuit and the underlying logical circuit and handle the qubit mapping problem as if on a larger interconnected circuit. It will still be important to determine the optimal routing path of any two-qubit gates under the QECC mechanism, and we believe that our robust qubit mapping strategy on large-scale circuits will become imperative.}

\subsubsection{Target Topologies}\label{section:targetedTopologies}

{\color{black}Our experiments are based on} the architectures of a series of IBMQ machines including the 16-qubit Guadalupe, 27-qubit Kolkata, 65-qubit Brooklyn, 127-qubit Washington, 433-qubit Osprey and 1121-qubit Condor. Furthermore, to emulate devices beyond 1121-qubit, we adopt the heavy-hex pattern from the existing IBM machines \cite{ibm_ibm_2021-1} and extrapolate the topology up to a remarkable 11,969-qubit, allowing us to assess the scalability of our algorithm thoroughly. These emulated devices are 2,129, 3,457, 5,105, 7,073, 9,361, and 11,969 physical qubits. 

\begin{figure}
     \centering
     \begin{subfigure}[b]{0.45\textwidth}
         \centering
         \includegraphics[width=\textwidth]{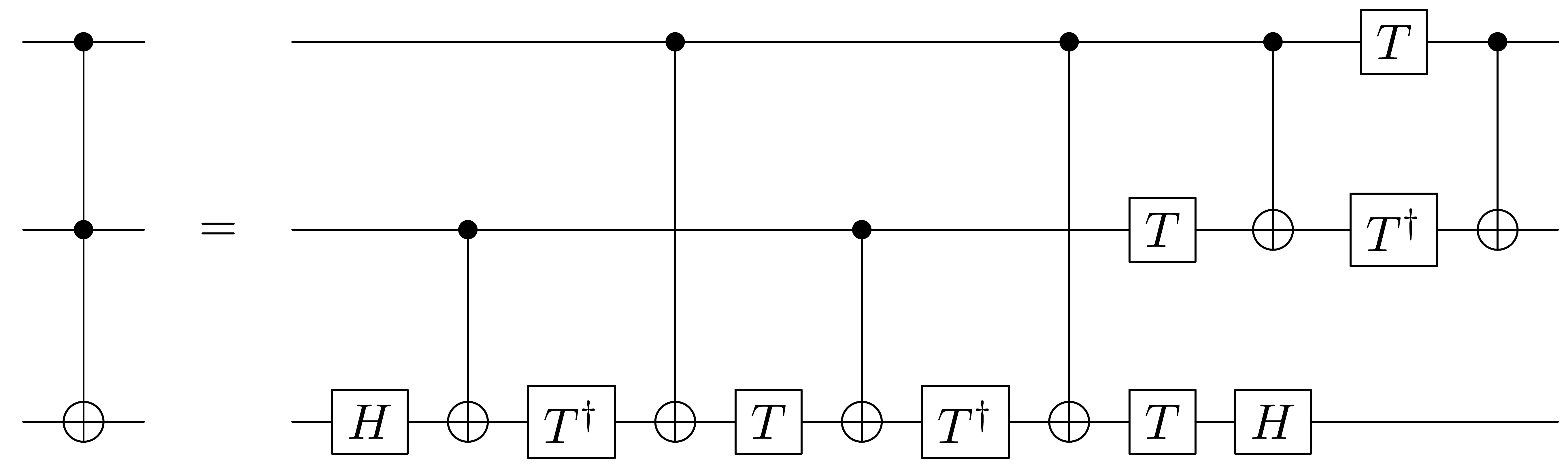}
         \caption{Toffoli \cite{nielsen_quantum_2010}}
         \label{fig:tofDecomp}
     \end{subfigure}
     \hfill
     \begin{subfigure}[b]{0.45\textwidth}
         \centering
         \includegraphics[width=\textwidth]{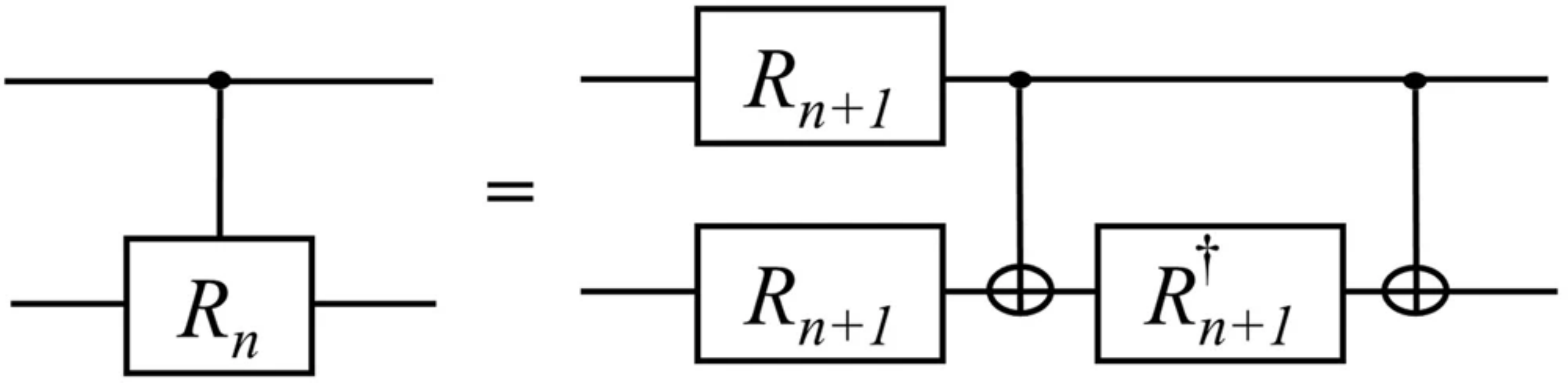}
         \caption{Control rotation \cite{kim_efficient_2018}. $R_n$ represents a rotate-z gate with phase $1/2^n$}
         
     \end{subfigure}
     \caption{Gate Decomposition}
     \label{fig:gateDecomp}
\end{figure}

There are two limitations, imposed by the characteristics of the IBMQ machines, to our experiments. First, with the designated qubit counts of the machines, we have to select the smallest available device that can accommodate the size of the circuit. For example, for the circuit with 256 logical qubits, we will map it to the topology with 433 physical qubits. Second, since IBMQ does not support control-rotation and Toffoli gates, we have to decompose these complex gates into their fundamentally-equivalent counterparts, as depicted in Fig.~\ref{fig:gateDecomp}.

\subsubsection{Benchmarks}
To assess the different superior aspects of our framework, including performance, scalability, and futurism, we employ distinct sets of benchmarks in our evaluations. 

Firstly, for the performance assessment, we focus on the circuits that are big enough so that the optimum solution can not be acquired by the exact approach such as OLSQ2 \cite{lin_scalable_2023}, and are smaller enough so that the compared methods (c.f. Section~\ref{section:comparedMethod}) can all generate meaningful results with heuristics. The chosen benchmark consists of circuits proposed in SABRE-large \cite{li_tackling_2019}, along with its extended form as presented by Wille et al. {\color{black}(QMAP\footnote{https://github.com/cda-tum/mqt-qmap/tree/main/examples})} \cite{wille_mqt_2023}. These circuits constitute a subset\footnote{Our focus for performance evaluation is on circuits with Ideal Circuit Costs (c.f.~\emph{Definition~\ref{def:cost}}) surpassing 500. For those with Ideal Circuit Cost smaller than 500, since it is often that their optimum mapping results can be achieved by exact methods, we will skip them for performance comparison and yet include them in the initial placement experiments (Section~\ref{section:initialPlacementExp}).} of RevLib \cite{wille_revlib_2008} and have been decomposed into single- and double-qubit gates. They are up to 185,000 gates and are no more than 16 qubits. 

Secondly, for our scalability experiments, we either concentrate on the circuits with the scalable and repeatable architectures, such as Quantum Fourier Transform (QFT), Galois Field (GF), and Inst, or on the benchmark that contains large oracle circuits such as ISCAS’85 \cite{brglez_neutral_1985}. The QFT circuits are generated using the method outlined in \cite{nielsen_quantum_2010}, and Galois Field and Inst circuits are respectively sourced from \cite{kissinger_pyzx_2020} and \cite{chen_veriqbench_2022}. ISCAS’85, on the other hand, is a collection of classical EDA circuits \cite{brglez_neutral_1985}. To prepare the quantum version of the ISCAS’85 benchmark, we initially transform the circuits into XOR-Majority Graph (XMG) form utilizing the ABC tool \cite{group_abc_2007}. This transformation enables a direct mapping from the traditional AND, OR, and NOT gates to the X, CX, and Toffoli gates in a quantum circuit. Subsequently, we decompose the Toffoli gates into a combination of single- and double-qubit gates as Fig.~\ref{fig:tofDecomp}. 

Lastly, to explore the futurism of quantum computing beyond the NISQ era, we conduct experiments on the emulated large-scale quantum circuits as mentioned in Section~\ref{section:targetedTopologies}.

\subsection{Experiments on Initial Placement}\label{section:initialPlacementExp}

\begin{figure}
    \centering
    \includegraphics[width=0.7\textwidth]{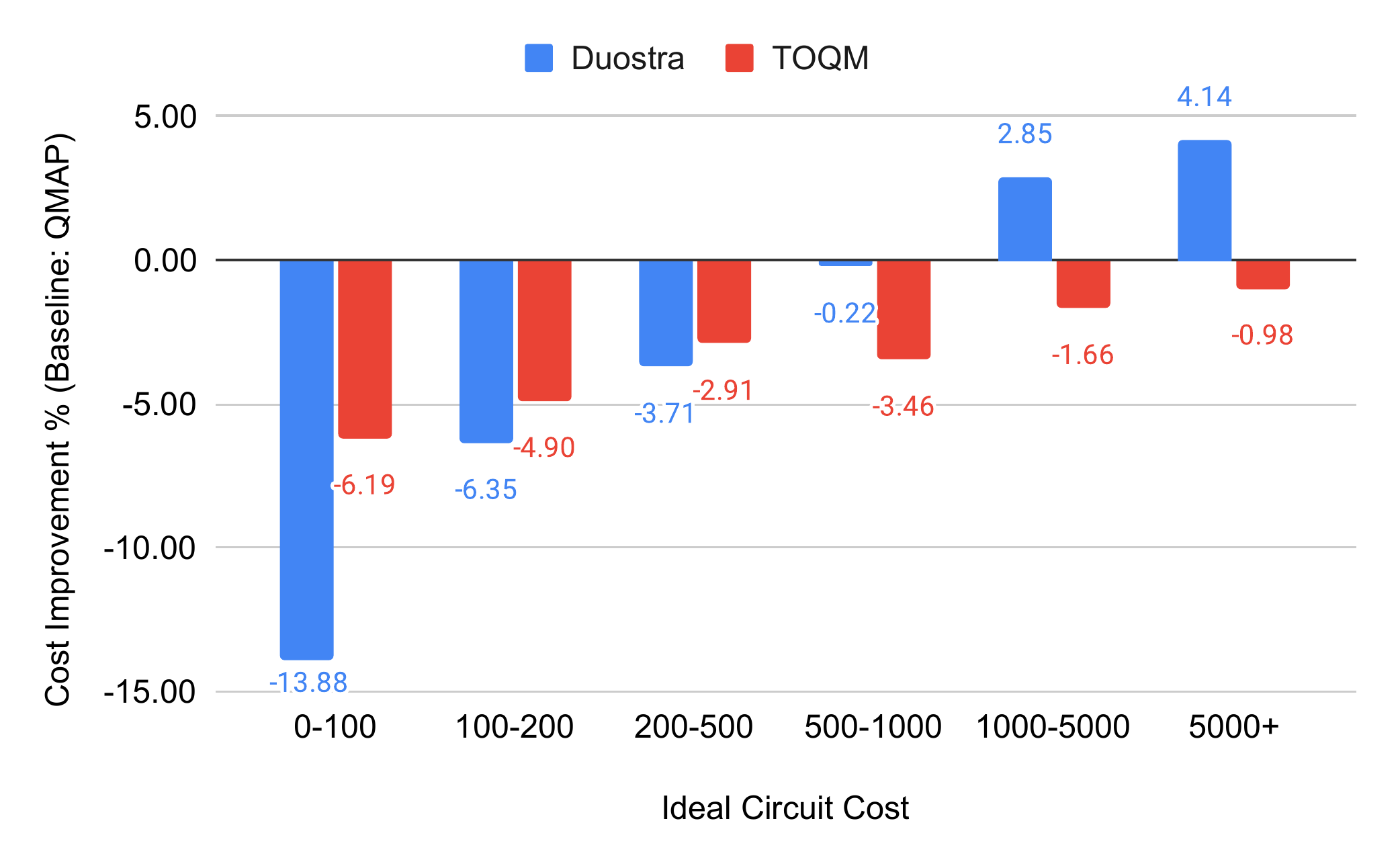}
    \caption{Mapping result on all benchmarks in QMAP \cite{wille_mqt_2023}. With the ideal cost growth, the improvement of Duostra grows.}
    \label{fig:initial}
\end{figure}

As mentioned in Section~\ref{section:initialPlacement}, to testify to the impact of initial placement on the qubit mapping problem, we compare with the QMAP method, which incorporates an adaptive initial placement heuristic for better mapping results \cite{peham_optimal_2022}, on various sizes of circuits. The experimental results agree with our speculations – As the circuit depth increases along with its gate count, the Mapping Cost introduced by the scheduling and routing procedures starts to dominate the overall cost. Therefore, the impact of the initial placement becomes insignificant.

For smaller circuits whose Ideal Circuit Costs (c.f.~\emph{Definition~\ref{def:cost}}) are smaller than 500, QMAP’s effort in exploring better initial placement pays back as it exhibits superior performance over the Duostra LE Search and TOQM. However, as the Ideal Circuit Cost grows, Duostra turns around and outperforms QMAP (Fig.~\ref{fig:initial}). 

It is noteworthy that achieving optimal qubit mapping results on small circuits (Ideal Circuit Cost~<~500) is not our focus in this paper because the exact solution naturally overshadows heuristic approaches for circuits of this size. Thus, the value of heuristic-based methods shines through when the exact solution is not feasible. As indicated in Fig.~\ref{fig:initial}, our approach prevails as the circuit size increases, which is our main contribution.

\subsection{Performance Evaluation on Mid-Size Circuits}

To evaluate the performance of various heuristics algorithms of different tools, we conduct experiments on the mid-size circuits adopted from the SABRE-large test suite \cite{li_tackling_2019} and its extended set in \cite{wille_mqt_2023}. Note that we filter out the small circuits whose Ideal Circuit Costs are smaller than 500 because their optimum mapping solutions can be attained within one day by the exact approaches such as OLSQ \cite{tan_optimal_2021, lin_scalable_2023}. The selected circuits are of sizes up to 185,000 gates and less than 16 qubits. Consequently, all the compared tools can finish the qubit mapping task for every circuit within one minute, allowing the full exploitation of the heuristic algorithms of the tools.

\begin{table}
\scriptsize
    \centering
    \caption{Performance on {\color{black}the SABRE-large} benchmark, with QMAP the baseline. {\color{black}The cost improvement is calculated as $\Delta (\%) = \frac{Cost_{QMAP}-Cost}{Cost_{QMAP}}$. Duostra LE is the only approach outperforming the baseline QMAP on average.}}
    \begin{tabular}{lrrrrrrrrrrrrr}
\toprule
    \multicolumn{1}{c}{\multirow{3}{*}{Circuit}}&\multicolumn{1}{c}{\multirow{3}{*}{$\#$Gate}}&\multicolumn{1}{c}{Ideal}&\multicolumn{1}{c}{QMAP}&\multicolumn{2}{c}{Qiskit \cite{aleksandrowicz_qiskit_2019}}&\multicolumn{2}{c}{TOQM \cite{zhang_time-optimal_2021}}&\multicolumn{2}{c}{\textcolor{black}{t$|ket\rangle$ \cite{sivarajah_tket_2021}}}&\multicolumn{2}{c}{\textcolor{black}{SABRE \cite{li_tackling_2019}}}&\multicolumn{2}{c}{Duostra LE}\\
\cmidrule{5-6}\cmidrule{7-8}\cmidrule{9-10}\cmidrule{11-12}\cmidrule{13-14}
 & &\multicolumn{1}{c}{Cost} &\multicolumn{1}{c}{\cite{wille_mqt_2023}*} & \multicolumn{1}{c}{Cost} & \multicolumn{1}{c}{$\Delta$($\%$)} & \multicolumn{1}{c}{Cost} & \multicolumn{1}{c}{$\Delta$($\%$)} & \multicolumn{1}{c}{\textcolor{black}{Cost}}  & \multicolumn{1}{c}{\textcolor{black}{$\Delta$($\%$)}} & \multicolumn{1}{c}{\textcolor{black}{Cost}} & \multicolumn{1}{c}{\textcolor{black}{$\Delta$($\%$)}} & \multicolumn{1}{c}{Cost}& \multicolumn{1}{c}{$\Delta$($\%$)} \\
\midrule
\midrule
cm82a & 650 & 571 & 1485 & 1803 & -21.4 & 1554 & -4.7 & \textcolor{black}{1488} & \textcolor{black}{-0.2} & \textcolor{black}{1806} & \textcolor{black}{-21.6} & 1495 & -0.7 \\ 
        rd53\_251 & 1291 & 1203 & 3154 & 3712 & -17.7 & 3348 & -6.2 & \textcolor{black}{3316} & \textcolor{black}{-5.1} & \textcolor{black}{3814} & \textcolor{black}{-20.9} & 3192 & -1.2 \\ 
        cm42a & 1776 & 1574 & 4515 & 5128 & -13.6 & 4472 & 1.0 & \textcolor{black}{4850} & \textcolor{black}{-7.4} & \textcolor{black}{5426} & \textcolor{black}{-20.2} & 4245 & 6.0 \\ 
        pm1\_249 & 1776 & 1574 & 4515 & 5140 & -13.8 & 4472 & 1.0 & \textcolor{black}{4850} & \textcolor{black}{-7.4} & \textcolor{black}{5426} & \textcolor{black}{-20.2} & 4245 & 6.0 \\ 
        sqrt8\_260 & 3009 & 2779 & 8078 & 9211 & -14.0 & 7863 & 2.7 & \textcolor{black}{8902} & \textcolor{black}{-10.2} & \textcolor{black}{9420} & \textcolor{black}{-16.6} & 7561 & 6.4 \\ 
        z4\_268 & 3073 & 2756 & 7793 & 9165 & -17.6 & 7887 & -1.2 & \textcolor{black}{8651} & \textcolor{black}{-11.0} & \textcolor{black}{9576} & \textcolor{black}{-22.9} & 7441 & 4.5 \\ 
        adr4\_197 & 3439 & 3088 & 8643 & 10372 & -20.0 & 8859 & -2.5 & \textcolor{black}{9721} & \textcolor{black}{-12.5} & \textcolor{black}{10627} & \textcolor{black}{-23.0} & 8225 & 4.8 \\ 
        rd73\_252 & 5321 & 4829 & 13350 & 15817 & -18.5 & 13869 & -3.9 & \textcolor{black}{17884} & \textcolor{black}{-34.0} & \textcolor{black}{16865} & \textcolor{black}{-26.3} & 12895 & 3.4 \\ 
        cycle10 & 6050 & 5662 & 15897 & 18682 & -17.5 & 15880 & 0.1 & \textcolor{black}{18997} & \textcolor{black}{-19.5} & \textcolor{black}{19193} & \textcolor{black}{-20.7} & 15321 & 3.6 \\ 
        sqrt\_7 & 7630 & 6367 & 18930 & 21086 & -11.4 & 18049 & 4.7 & \textcolor{black}{22055} & \textcolor{black}{-16.5} & \textcolor{black}{21316} & \textcolor{black}{-12.6} & 17621 & 6.9 \\ 
        ham15 & 8763 & 8092 & 22387 & 26201 & -17.0 & 23048 & -3.0 & \textcolor{black}{26644} & \textcolor{black}{-19.0} & \textcolor{black}{27367} & \textcolor{black}{-22.3} & 21652 & 3.3 \\ 
        dc2\_222 & 9462 & 8759 & 25256 & 28756 & -13.9 & 24603 & 2.6 & \textcolor{black}{28337} & \textcolor{black}{-12.2} & \textcolor{black}{29886} & \textcolor{black}{-18.3} & 23773 & 5.9 \\ 
        sqn\_258 & 10223 & 9176 & 26850 & 30296 & -12.8 & 26502 & 1.3 & \textcolor{black}{29694} & \textcolor{black}{-10.6} & \textcolor{black}{32079} & \textcolor{black}{-19.5} & 24496 & 8.8 \\ 
        inc\_237 & 10619 & 9790 & 28281 & 31963 & -13.0 & 27256 & 3.6 & \textcolor{black}{30512} & \textcolor{black}{-7.9} & \textcolor{black}{33075} & \textcolor{black}{-17.0} & 26500 & 6.3 \\ 
        cm85a & 11414 & 10630 & 30715 & 35015 & -14.0 & 30157 & 1.8 & \textcolor{black}{35617} & \textcolor{black}{-16.0} & \textcolor{black}{36385} & \textcolor{black}{-18.5} & 29076 & 5.3 \\ 
        rd84\_253 & 13658 & 12176 & 35874 & 40567 & -13.1 & 34876 & 2.8 & \textcolor{black}{40122} & \textcolor{black}{-11.8} & \textcolor{black}{42543} & \textcolor{black}{-18.6} & 33019 & 8.0 \\ 
        root\_255 & 17159 & 14799 & 43604 & 49606 & -13.8 & 42383 & 2.8 & \textcolor{black}{47990} & \textcolor{black}{-10.1} & \textcolor{black}{52463} & \textcolor{black}{-20.3} & 40671 & 6.7 \\ 
        mlp4\_245 & 18852 & 17258 & 49755 & 57379 & -15.3 & 48990 & 1.5 & \textcolor{black}{55141} & \textcolor{black}{-10.8} & \textcolor{black}{59236} & \textcolor{black}{-19.1} & 47199 & 5.1 \\ 
        urf2\_277 & 20112 & 19698 & 55709 & 66283 & -19.0 & 59989 & -7.7 & \textcolor{black}{56785} & \textcolor{black}{-1.9} & \textcolor{black}{69332} & \textcolor{black}{-24.5} & 55126 & 1.0 \\ 
        life\_238 & 22445 & 20867 & 60967 & 68673 & -12.6 & 58932 & 3.3 & \textcolor{black}{65319} & \textcolor{black}{-7.1} & \textcolor{black}{71679} & \textcolor{black}{-17.6} & 56546 & 7.3 \\ 
        9symml & 34881 & 32084 & 94360 & 106104 & -12.5 & 90976 & 3.6 & \textcolor{black}{100672} & \textcolor{black}{-6.7} & \textcolor{black}{111521} & \textcolor{black}{-18.2} & 86615 & 8.2 \\ 
        dist\_223 & 38046 & 32968 & 97097 & 110229 & -13.5 & 95564 & 1.6 & \textcolor{black}{107263} & \textcolor{black}{-10.5} & \textcolor{black}{116529} & \textcolor{black}{-20.0} & 90288 & 7.0 \\ 
        urf1\_278 & 54766 & 53256 & 152057 & 183630 & -20.8 & 160414 & -5.5 & \textcolor{black}{185152} & \textcolor{black}{-21.8} & \textcolor{black}{194702} & \textcolor{black}{-28.1} & 147585 & 2.9 \\ 
        hwb8\_113 & 69380 & 64758 & 175618 & 210658 & -20.0 & 183657 & -4.6 & \textcolor{black}{184046} & \textcolor{black}{-4.8} & \textcolor{black}{220665} & \textcolor{black}{-25.7} & 172575 & 1.7 \\ 
        urf1\_149 & 184864 & 172518 & 413994 & 534114 & -29.0 & 466086 & -12.6 & \textcolor{black}{499771} & \textcolor{black}{-20.7} & \textcolor{black}{572947} & \textcolor{black}{-38.4} & 437037 & -5.6 \\
     \midrule
     \multicolumn{4}{l}{Average Cost Improvement}     &  & -16.2 & & -0.70 && \textcolor{black}{-11.8} && \textcolor{black}{-21.2} & & 4.5\\
\bottomrule
\end{tabular}
    \label{tab:SABRE}
\end{table}

The experimental results are presented in Table~\ref{tab:SABRE}. {\color{black}Columns 2 and 3 list the number of gates and the Ideal Circuit Costs of the benchmark circuits, respectively.} We choose QMAP \cite{wille_mqt_2023} as the baseline method (note: marked with ``*'') because it performs fairly well on these testcases. {\color{black} Columns 5 to 12 in the table report the Mapping Costs and the comparison with QMAP of IBM Qiskit, TOQM, t$|ket\rangle$, and SABRE. The table indicate that on average, IBM Qiskit, t$|ket\rangle$, and SABRE exhibit inferior performance compared to QMAP by 16.2\%, 11.8\% and 21.2\% while TOQM demonstrates relatively competitive results with only 0.7\% difference on average.}

The qubit mapping results of our algorithm are listed in the last two columns. For the mid-size circuits, we employ the LE Search approach with the depth set to 4. It shows that we can outperform QMAP for the average of {\color{black}$4.5\%$} as we win 22 of the 25 testcases. 

\begin{figure}
    \centering
    \includegraphics[width=\linewidth]{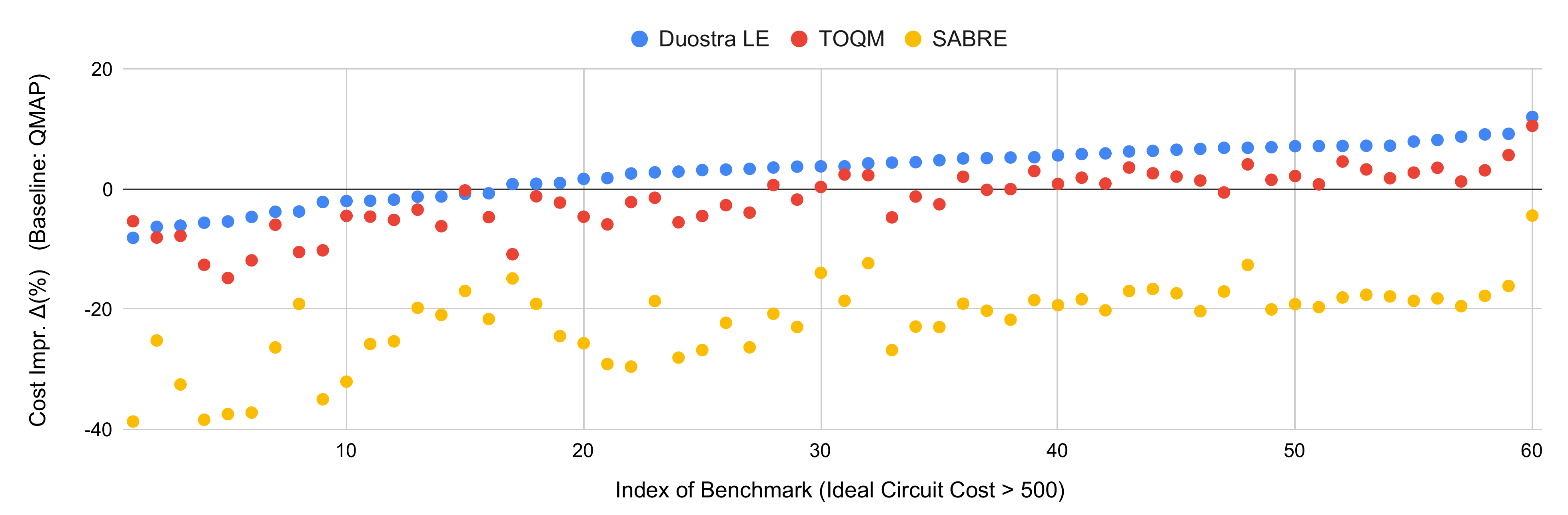}
    \includegraphics[width=\linewidth]{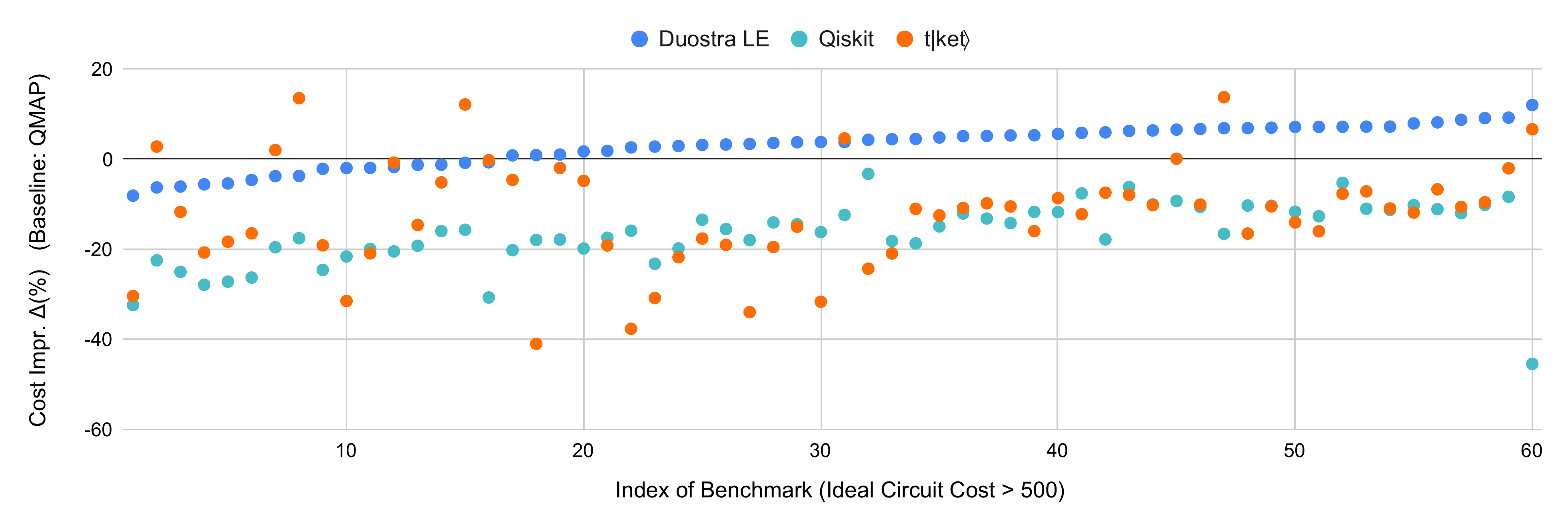}
    \caption[]{Performance on the extended benchmark provided by QMAP \cite{wille_mqt_2023}. Only the circuits with the Initial Circuit Cost greater than 500 are included in the experiment. We select QMAP's result selected as the baseline. The indices are sorted by the relative improvements of the Duostra LE search. Duostra LE search outperforms the others in most cases.}
    \label{fig:QMAP}
\end{figure}

We further conduct the experiments on the extended set of the benchmark circuits \cite{wille_mqt_2023}. As shown in Fig.~\ref{fig:QMAP}, our algorithm can top the other tools on {\color{black}70.0\% (42/60)} of the cases and improve over QMAP for the average of 4\% in Mapping Costs.

\begin{table}[t]
\caption{{\color{black}Performance on various large benchmarks. The highlighted results represent the best one among QMAP, Qiskit, t$|ket\rangle$, and SABRE. The cost improvement is calculated based on the best results among the counterparts ($\Delta (\%) = \frac{Cost_{best}-Cost}{Cost_{best}}$). TLE stands for exceeding the time limit one day. Duostra SP estimation outperforms the others in most cases.}}
\tiny
    \centering
    {\color{black}\begin{tabular}{rrrrrrrrrrr|rrr}
\toprule
    \multicolumn{1}{c}{\multirow{3}{*}{Circuit}}&\multicolumn{1}{c}{Cir.}&\multicolumn{1}{c}{Top.}&\multicolumn{2}{c}{QMAP \cite{wille_mqt_2023}}&\multicolumn{2}{c}{Qiskit \cite{aleksandrowicz_qiskit_2019}}&\multicolumn{2}{c}{t$|ket\rangle$ \cite{sivarajah_tket_2021}}&\multicolumn{2}{c}{SABRE \cite{li_tackling_2019}}&\multicolumn{3}{|c}{Duostra SP}\\
\cmidrule{4-5}\cmidrule{6-7}\cmidrule{8-9}\cmidrule{10-11}\cmidrule{12-14}
   & \multicolumn{1}{c}{$\#$Q}& \multicolumn{1}{c}{$\#$Q} & \multicolumn{1}{c}{Cost} & \multicolumn{1}{c}{Time}  & \multicolumn{1}{c}{Cost} & \multicolumn{1}{c}{Time} & \multicolumn{1}{c}{Cost} & \multicolumn{1}{c}{Time} &\multicolumn{1}{c}{Cost} & \multicolumn{1}{c}{Time} & \multicolumn{1}{|c}{Cost} & \multicolumn{1}{c}{Time} &\multicolumn{1}{c}{$\Delta (\%)$}\\
  \midrule
  \midrule
    \multirow{5}{*}{qft} & 2129 & 2129 & 23,754,260 & 57082 & TLE & TLE & TLE & TLE & \textbf{282,338} & 128.0 & 345,371 & 75.03 & -22.33 \\ 
        ~ & 1121 & 1121 & 5,649,206 & 3374 & 277,528 & 37460 & TLE & TLE & \textbf{127,537} & 28.07 & 178,639 & 11.32 & -40.07 \\ 
        ~ & 433 & 433 & 496,032 & 98.72 & 90,358 & 1000 & 255,807 & 1854 & \textbf{44,270} & 3.626 & 62,469 & 1.135 & -41.11 \\ 
        ~ & 127 & 127 & 62,668 & 3.083 & 20,026 & 12.12 & 29,942 & 25.70 & \textbf{11,212} & 0.254 & 14,436 & 0.242 & -28.75 \\ 
        ~ & 65 & 65 & 11,364 & 0.553 & 6,949 & 1.738 & 6,919 & 1.866 & \textbf{4,549} & 0.069 & 6,340 & 0.165 & -39.37 \\ \hline
        \multirow{5}{*}{\shortstack[1]{gf \cite{kissinger_pyzx_2020}}} & 768 & 1121 & 1,595,176 & 131.2 & \textbf{210,529} & 13758 & 1,647,228 & 11852 & 276,294 & 21.27 & 137,600 & 4.797 & 34.64 \\ 
        ~ & 384 & 433 & 375,442 & 21.57 & \textbf{87,893} & 654.1 & 393,352 & 906.8 & 102,912 & 5.407 & 63,848 & 0.871 & 27.36 \\ 
        ~ & 192 & 433 & 69,094 & 5.332 & \textbf{36,278} & 95.10 & 138,102 & 79.53 & 41,825 & 1.352 & 26,907 & 0.543 & 25.83 \\ 
        ~ & 96 & 127 & 23,243 & 1.086 & 16,156 & 6.791 & 28,644 & 7.239 & \textbf{15,643} & 0.333 & 11,669 & 0.190 & 25.40 \\ 
        ~ & 48 & 65 & \textbf{4,867} & 0.238 & 5,029 & 0.818 & 9,299 & 1.331 & 4,943 & 0.110 & 4,158 & 0.175 & 14.57 \\ \hline
        \multirow{5}{*}{\shortstack[1]{inst \cite{chen_veriqbench_2022}}} & 144 & 433 & 2,091 & 19.90 & 1,102 & 2.642 & 1,506 & 10.99 & \textbf{752} & 0.041 & 527 & 0.460 & 29.92 \\ 
        ~ & 100 & 127 & 1,482 & 1.727 & 788 & 0.749 & 596 & 1.941 & \textbf{569} & 0.021 & 399 & 0.180 & 29.88 \\ 
        ~ & 81 & 127 & 1,194 & 0.183 & 776 & 0.323 & 1,223 & 2.471 & \textbf{639} & 0.038 & 416 & 0.180 & 34.90 \\ 
        ~ & 64 & 65 & 877 & 0.106 & \textbf{441} & 0.104 & 849 & 2.994 & 448 & 0.028 & 347 & 0.160 & 21.32 \\ 
        ~ & 49 & 65 & 634 & 0.069 & 442 & 0.096 & 456 & 9.176 & \textbf{418} & 0.016 & 281 & 0.180 & 32.78 \\ \hline
          \multirow{9}{*}{\shortstack[1]{ISCAS’85 \cite{brglez_neutral_1985}}} & 1573 & 2129 & TLE & TLE & 51,768 & 646.7 & 546,242 & 17895 & \textbf{37,518} & 1.366 & 12,297 & 24.49 & 67.22 \\
        ~ & 1380 & 2129 & TLE & TLE & 92,227 & 494.0 & 416,778 & 16827 & \textbf{79,323} & 1.167 & 15,006 & 27.69 & 81.08 \\ 
        ~ & 991 & 1121 & TLE & TLE & 50,894 & 143.6 & 245,659 & 4011 & \textbf{41,505} & 0.794 & 14,154 & 4.310 & 65.90 \\ 
        ~ & 840 & 1121 & TLE & TLE & 45,947 & 189.9 & 197,757 & 2553 & \textbf{31,431} & 0.861 & 13,545 & 3.340 & 56.91 \\ 
        ~ & 660 & 1121 & TLE & TLE & 20,045 & 70.83 & 131,372 & 885.8 & \textbf{15,164} & 0.385 & 6,199 & 5.090 & 59.12 \\
        ~ & 255 & 433 & \textbf{10,313} & 4.580 & 12,874 & 7.829 & 20,506 & 29.01 & 11,127 & 0.139 & 4,850 & 0.480 & 52.97 \\ 
        ~ & 202 & 433 & \textbf{6,807} & 9.186 & 7,915 & 6.041 & 14,673 & 22.30 & 7,566 & 0.102 & 3,612 & 0.460 & 46.94 \\ 
        ~ & 188 & 433 & \textbf{6,587} & 41.75 & 8,821 & 5.972 & 11,755 & 12.41 & 7,023 & 0.081 & 3,705 & 0.440 & 43.75 \\
        ~ & 172 & 433 & \textbf{7,322} & 552.7 & 9,269 & 6.58 & 17,278 & 18.97 & 7,550 & 0.111 & 4,630 & 0.460 & 36.77 \\ 
        \hline 
        \multirow{3}{*}{adder \cite{li_qasmbench_2022}} & 433 & 433 & \textbf{14,559} & 6.788 & 16,228 & 42.95 & 30,906 & 150.3 & 17,693 & 0.255 & 8,961 & 0.450 & 38.45 \\ 
        ~ & 118 & 127 & 5,048 & 0.175 & 4,717 & 2.899 & 5,542 & 5.399 & \textbf{3,096} & 0.099 & 2,694 & 0.930 & 12.98 \\ 
        ~ & 64 & 65 & 1,747 & 0.066 & 2,492 & 0.851 & 2,482 & 1.725 & \textbf{1,684} & 0.024 & 1,670 & 0.170 & 0.83 \\\hline 
         \multirow{1}{*}{dnn \cite{li_qasmbench_2022}} & 51 & 65 & 1,346 & 0.065 & 1,935 & 0.745 & \textbf{1,166} & 0.791 & 1,541 & 0.124 & 1,036 & 0.160 & 11.15 \\ \hline
        \multirow{3}{*}{knn \cite{li_qasmbench_2022}} & 341 & 433 & 9,664 & 0.729 & 11,258 & 7.823 & 22,894 & 15.04 & \textbf{9,127} & 0.114 & 6,510 & 0.470 & 28.67 \\ 
        ~ & 129 & 433 & 3,518 & 1.058 & 3,750 & 2.965 & 8,317 & 2.754 & \textbf{3,170} & 0.103 & 2,467 & 0.420 & 22.18 \\ 
        ~ & 67 & 127 & 1,622 & 0.069 & 1,851 & 0.452 & 1,622 & 1.569 & \textbf{1,329} & 0.091 & 1,355 & 0.200 & -1.96 \\ \hline
        \multirow{4}{*}{qugan \cite{li_qasmbench_2022}} & 395 & 433 & \textbf{11,362} & 2.274 & 18,879 & 26.80 & 22,442 & 469.7 & 26,104 & 1.328 & 8,698 & 0.440 & 23.45 \\ 
        ~ & 111 & 127 & \textbf{3,193} & 0.204 & 4,850 & 1.896 & 3,792 & 5.788 & 4,270 & 0.063 & 2,369 & 0.200 & 25.81 \\ 
        ~ & 71 & 127 & \textbf{1,806} & 0.111 & 3,250 & 1.077 & 2,689 & 1.482 & 2,036 & 0.050 & 1,598 & 0.200 & 11.52 \\ 
        ~ & 39 & 65 & 1,051 & 0.042 & 1,694 & 0.305 & \textbf{883} & 0.408 & 892 & 0.031 & 786 & 0.170 & 10.99 \\ \hline
         \multirow{4}{*}{\shortstack[1]{multiplier \cite{li_qasmbench_2022}}} & 400 & 433 & TLE & TLE & 1,263,928 & 1251 & 2,505,716 & 9912 & \textbf{981,051} & 10.39 & 796,618 & 1.540 & 18.80 \\ 
        ~ & 350 & 433 & TLE & TLE & 963,091 & 964.5 & 1,904,264 & 6060 & \textbf{837,539} & 5.673 & 612,075 & 1.300 & 26.92 \\ 
        ~ & 75 & 127 & 36,413 & 2.419 & 42,943 & 10.35 & 43,254 & 11.50 & \textbf{26,762} & 0.247 & 28,284 & 0.190 & -5.69 \\ 
        ~ & 45 & 65 & \textbf{10,845} & 0.365 & 13,787 & 1.720 & 13,121 & 1.885 & 11,769 & 0.103 & 9,952 & 0.160 & 8.23 \\ 
        \midrule
        \midrule
        \multicolumn{3}{c}{Average Cost Improvement}&& ~ & ~ & ~ & ~ & ~ & ~ & ~ &  && 21.75 \\
      \bottomrule
      \end{tabular}}
    \vspace{10px}
    \label{table:large-scale}
\end{table}

\subsection{Scalability Experiment on Large Circuits}

\begin{figure}
     \centering
     \begin{subfigure}[b]{0.49\textwidth}
         \centering
         \includegraphics[width=\textwidth]{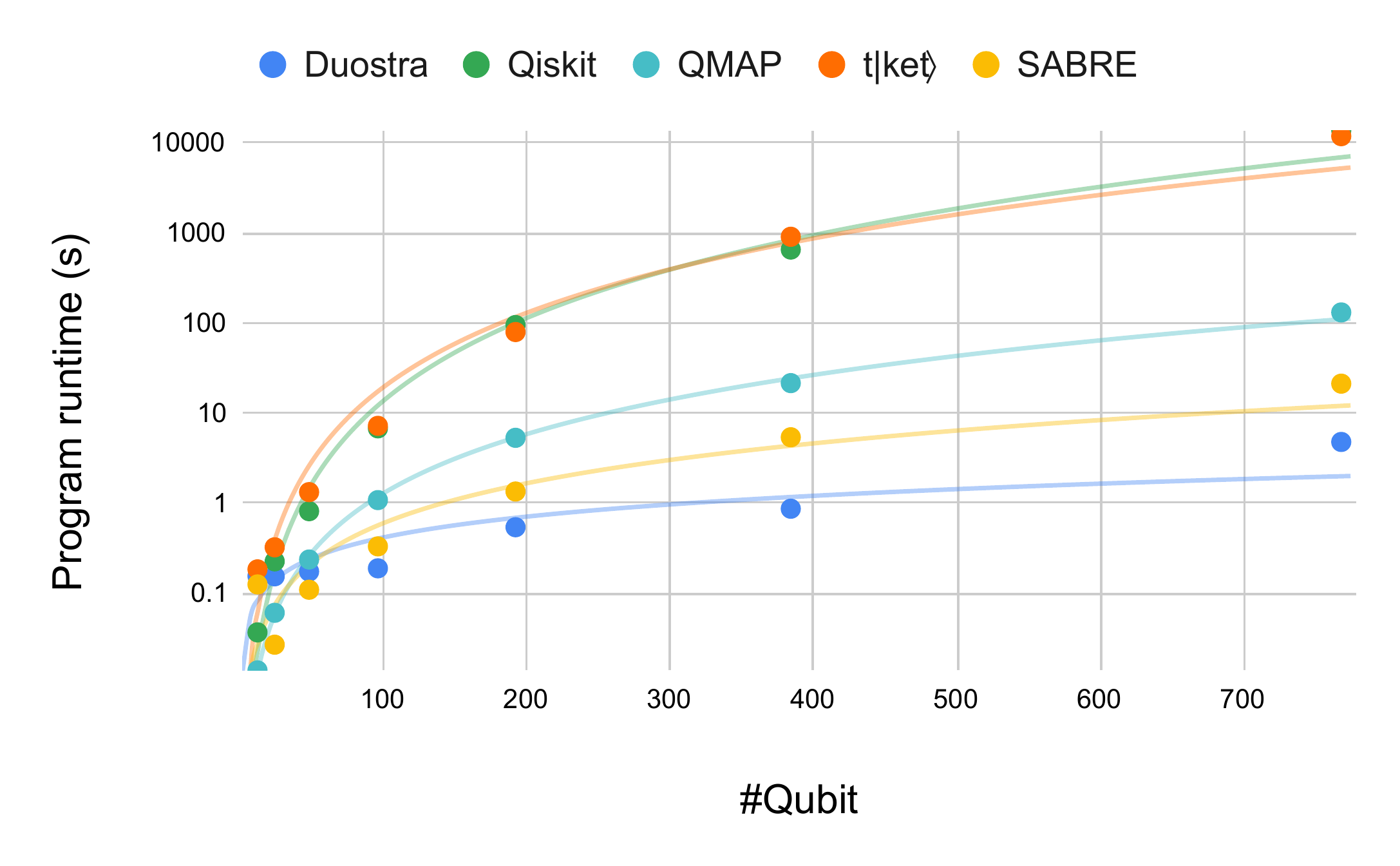}
        \caption{Program runtime}
        \label{fig: GF time}
         
     \end{subfigure}
     \hfill
     \begin{subfigure}[b]{0.49\textwidth}
         \centering
         \includegraphics[width=\textwidth]{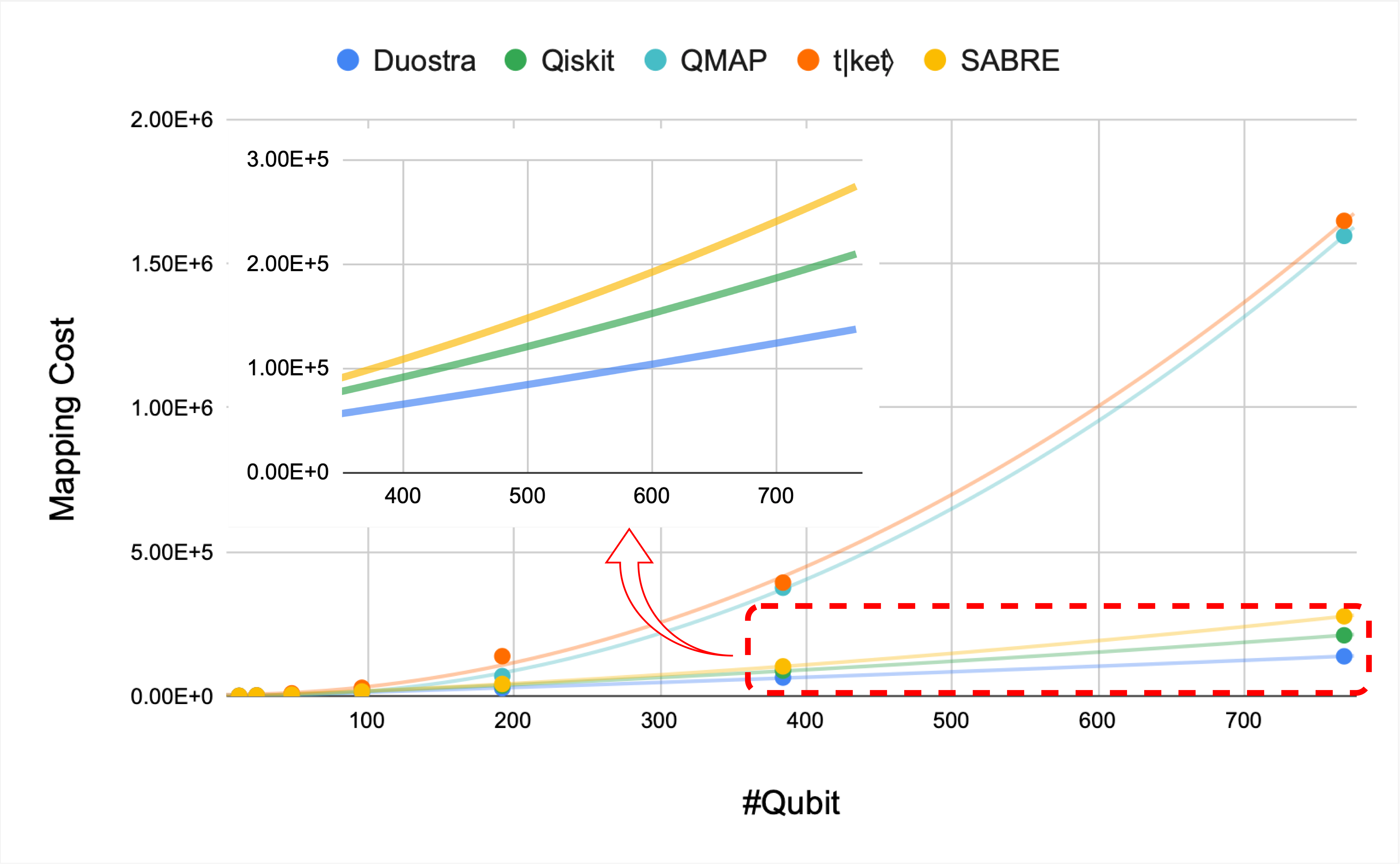}
        \caption{Mapping Cost}
        \label{fig: GF cost}
     \end{subfigure}
     \caption{Trending curves for the set of GF circuits.}
     \label{fig:GF}
\end{figure}

{\color{black}
To assess the scalability of our algorithm, we conducted experiments on circuits with scalable and repeatable architectures, including Quantum Fourier Transform (QFT), Galois Field (GF) \cite{kissinger_pyzx_2020}, and Inst \cite{chen_veriqbench_2022}. We implemented the QFT circuits for 65, 127, 433, 1121, and 2129 qubits, and decomposed them into circuits with basic single- and double-qubit gates \cite{nielsen_quantum_2010}. In addition, we included large oracle circuits such as ISCAS’85 \cite{brglez_neutral_1985} and a benchmark suited for the NISQ era \cite{li_qasmbench_2022}. 

The results are outlined in Table~\ref{table:large-scale}. The columns ``Cir. \#Q'' and ``Top. \#Q'' stand for the numbers of qubits for the benchmark circuits and the target topologies of the hardware devices (i.e. IBMQ machines). We compared our mapping results with QMAP \cite{wille_mqt_2023}, Qiskit \cite{aleksandrowicz_qiskit_2019}, t$|ket\rangle$ \cite{sivarajah_tket_2021}, and SABRE \cite{li_tackling_2019}. For the method TOQM \cite{zhang_time-optimal_2021} included in the mid-size circuit experiments, it failed to execute on the large-size circuits (TLE). Therefore, we exclude it for the comparison here. For our setup in this experiment, we employed SP Estimation instead of LE Search scheduler in order to trade off certain mapping optimality for better runtime performance.

It can be shown that the runtime performance of both Duostra and SABRE is superior to the other three, as the latter failed to finish certain cases within 1,000 seconds or even ran for more a day (marked as TLE). If we look closely into the numbers, we can see that Duostra is insignificantly slower than SABRE for the cases that run for less than 1 second, while its performance prevails for larger cases. This trend can be verified by the set of GF circuits as shown in Fig.~\ref{fig: GF time}. 

Regarding mapping results in Table~\ref{table:large-scale}, we highlight for each case the best one among the four compared methods, and compute our improvements with respect to these best results. The statistics show that we can achieve on the average 21.75\% improvement over the best performance of the compared methods. In general, due to the nature of the heuristic algorithms, different compared methods won at different benchmark sets — Qiskit excelled for larger GF circuits, QMAP performed better on qugan and some smaller ISCAS’85 benchmarks, and SABRE frequently produced the better results, particularly standing out for the QFT cases.

In short, the experimental results demonstrate the scalability and superiority of our qubit mapping framework for larger scale circuits. For the QFT cases that we lost to SABRE, it is likely due to the specifically repetitive patterns of the circuits. Our future work aims to enhance the Duostra Scheduler to identify and prioritize such patterns. We also plan to incorporate an automatic scheduler that can identify the suitable search depth, thereby striking a balance between solution quality and program runtime.}

\begin{figure}
     \centering
     \begin{subfigure}[b]{0.49\textwidth}
         \centering
         \includegraphics[width=\textwidth]{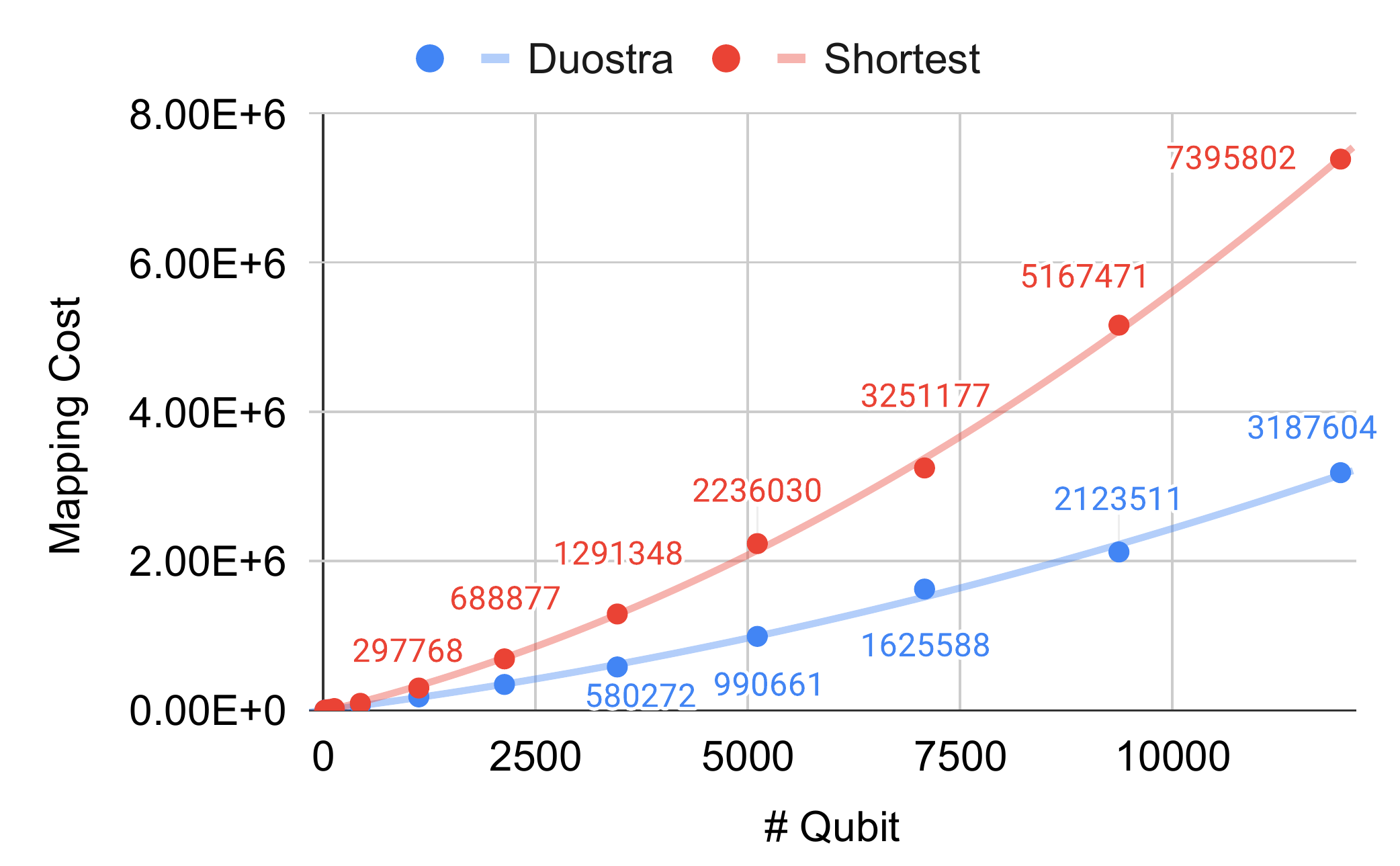}
         \caption{Mapping Cost}
         
     \end{subfigure}
     \hfill
     \begin{subfigure}[b]{0.49\textwidth}
         \centering
         \includegraphics[width=\textwidth]{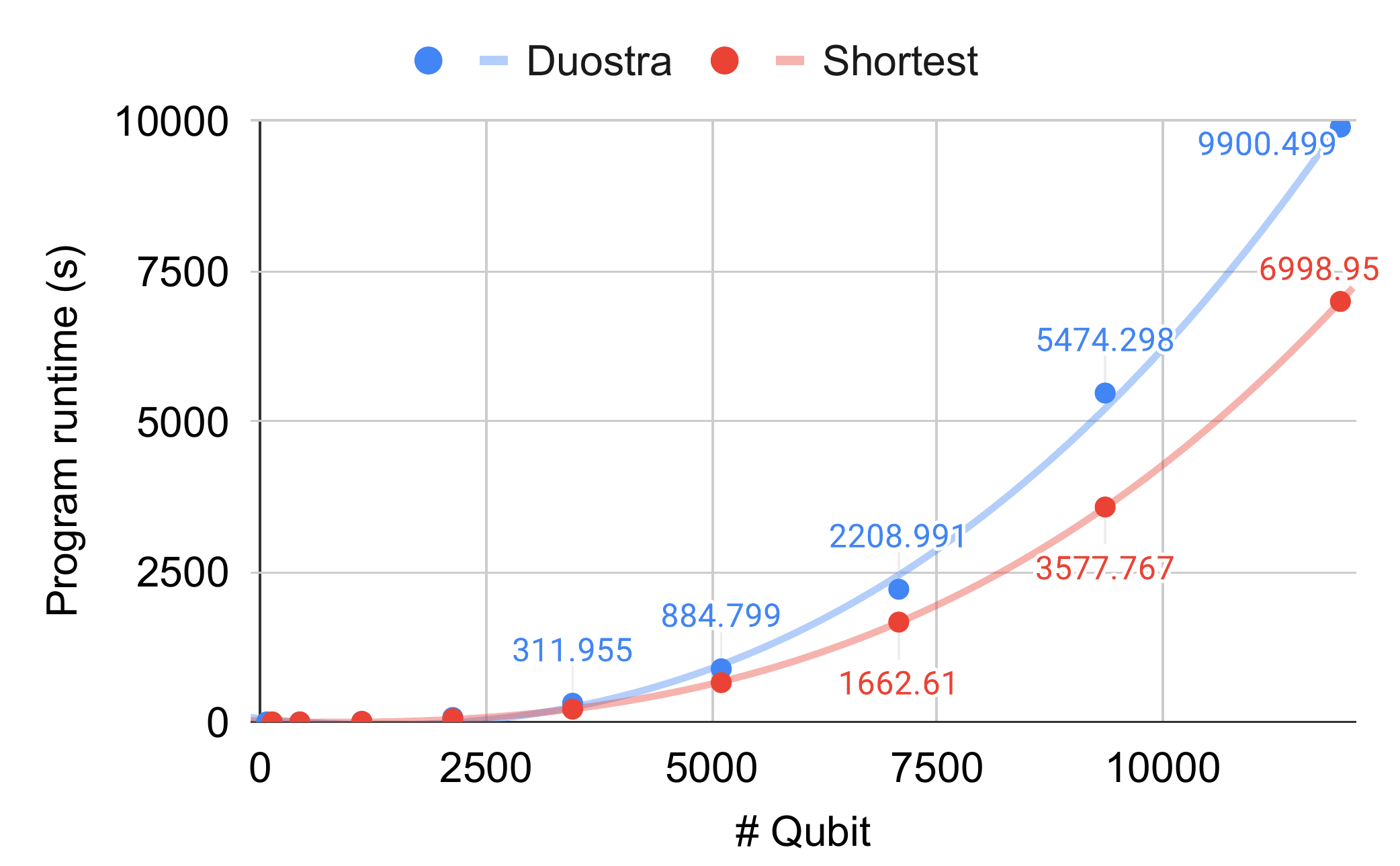}
         \caption{Program runtime}
         
     \end{subfigure}
     
     \begin{subfigure}[b]{0.49\textwidth}
         \centering
         \includegraphics[width=\textwidth]{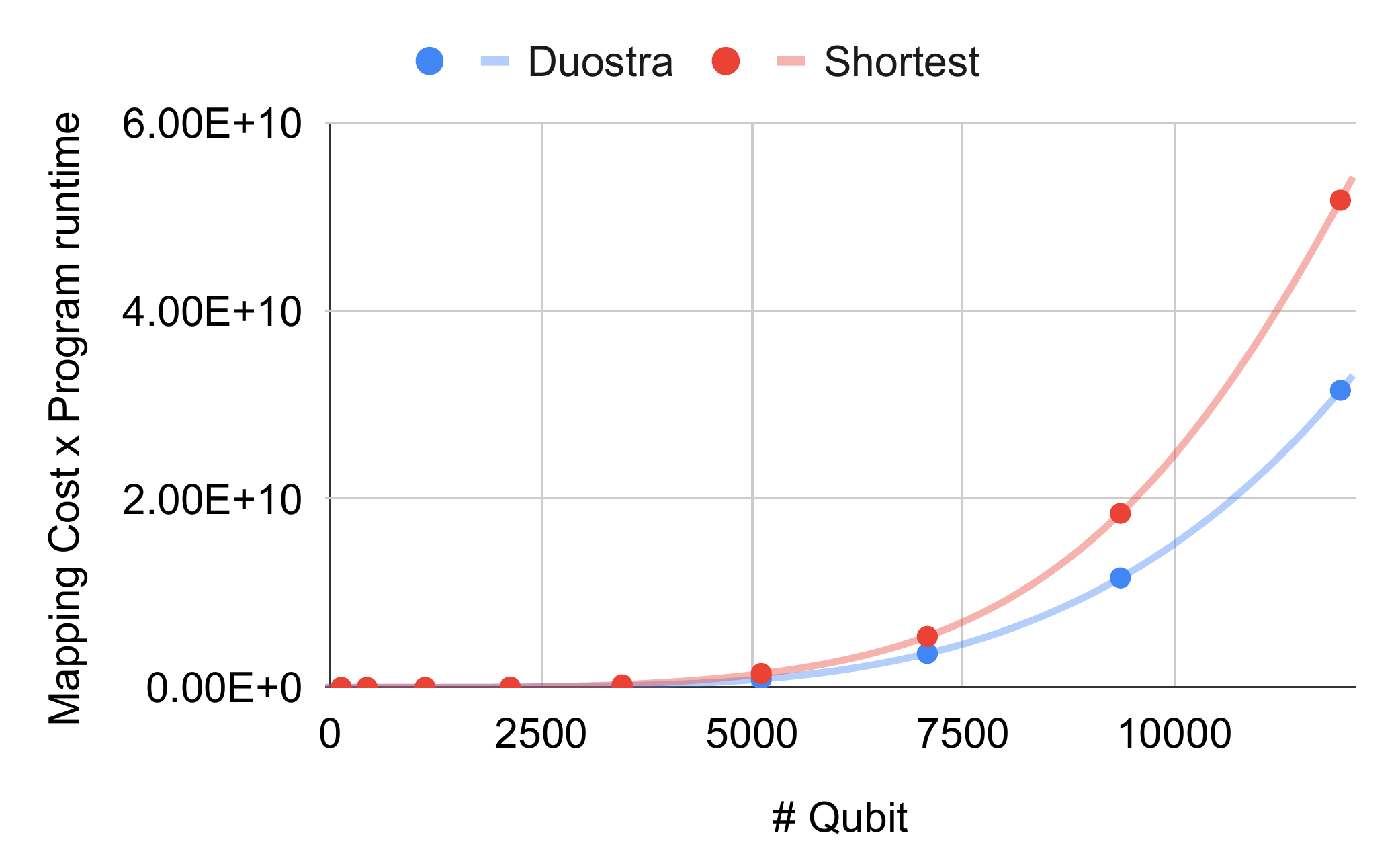}
         \caption{Mapping Cost x Program runtime}
         \label{fig:VeryLargeQFTsCT}
     \end{subfigure}
     \caption{Comparisons between Duostra and Shortest Path router on extra-large QFTs}
     \label{fig:VeryLargeQFTs}
\end{figure}

\subsection{Exploring the Futurism on Extra-Large Circuits}\label{section:Futurism}

To explore the futurism of quantum computing beyond the NISQ era, we conduct the qubit mapping experiments on QFT circuits up to 11,969 qubits. The target topologies are the emulated devices as described in Section \ref{section:targetedTopologies}. We choose the ``Shortest path'' approach (i.e. Shortest-Path router + SP Estimation scheduler) as the baseline comparison with ``Duostra'' to demonstrate the effect of  when considering the occupied time during the routing path selection. The selection of QFT circuits is attributed to its scalability potential and reputation as one of the most resource-demanding test sets, primarily due to the requirement for every qubit within the QFT to interact with all other qubits.

As depicted in Fig.~\ref{fig:VeryLargeQFTs}, the Mapping Cost associated with the Duostra router is consistently less than half of the cost generated by the Shortest Path router. Conversely, the runtime of the Shortest Path router is around two-thirds of that of Duostra. To facilitate a clearer and yet fair comparison, we multiply the cost and runtime values and plot them on Fig.~\ref{fig:VeryLargeQFTsCT}. The trend line of Duostra consistently appears lower, indicating that Duostra outperforms its counterpart considering both solution quality and efficiency as the circuit size increases. This leads us to conclude that, when compared to a simple approach like the Shortest Path router, which merely inserts SWAP gates to solve the problem, the Duostra Router's incorporation of additional information, such as the occupied time, contributes to its ability to generate superior results. 

\section{Conclusion} \label{section:CONCLUSION}

In this paper, we introduce a pioneering framework designed to address the qubit mapping challenge in large-scale quantum algorithms. Our framework encompasses an innovative Duostra router that can efficiently generate optimal routing paths for SWAP insertions on double-qubit gates, and a flexible scheduler that offers two heuristic scheduling algorithms, the Limitedly-Exhausitive (LE) Search and the Shortest-Path (SP) Estimation, for smaller- and larger-size circuits, respectively. The experimental results on various sets of benchmarks demonstrate the superiority of our framework on the qubit mapping problems. {\color{black}For instance, for the mid-size circuits (i.e. SABRE benchmark), we improve the Mapping Costs by 4.5\%, 5.2\%, 16.3\%, 20.7\% and 25.7\%, when compared to QMAP, TOQM, t$|ket\rangle$, Qiskit and SABRE, respectively. For the large benchmarks whose numbers of qubits exceed 50, our framework can achieve on the average 21.75\% improvements over the virtual best results among QMAP, t$|ket\rangle$, Qiskit and SABRE.} We also conduct experiments on extra-large QFT circuits up to 11,969 qubits. The results showcase our scalability to tackle the qubit mapping challenges for the quantum advantage.

{\color{black}Looking into the future, we plan to enhance our framework for broader applicability and more novel techniques. For example, fidelity issue is an inevitable factor for the practical quantum computation. Several studies have attempted to address the qubit and gate errors for the mapping problems \cite{das_case_2019, tannu_not_2019, murali_noise-adaptive_2019, li_qubits_2020, niu_hardware-aware_2020, das_adapt_2021, das_jigsaw_2021, jurcevic_demonstration_2021}, and some others have focused on scheduling gates to reduce the crosstalk errors \cite{mundada_suppression_2019, murali_software_2020, ding_systematic_2020, ohkura_simultaneous_2022}. In the near future, we will incorporate some of the error models and error correction circuits into our framework in order to answer the fault-tolerant qubit mapping problems.

On the other hand, several recent researches have explored innovated ideas to address the qubit mapping problems, including investigations into the potential of quantum teleportation~\cite{hillmich_exploiting_2021}, reinforcement learning techniques \cite{pozzi_using_2022}, and gate commutation and circuit transformation \cite{childs_circuit_2019, itoko_quantum_2019, itoko_optimization_2020, liu_not_2022}. However, scalability still remains a challenge for these methods due to their computational requirements. We plan to explore ways to integrate insights from these studies into our framework as the future work, aiming to achieve even greater efficiency and accuracy in qubit mapping for a broader range of quantum computing applications.}

\section*{Acknowledgement}
This work was partially supported by National Science and Technology Council of Taiwan under Grant NSTC 113-2119-M-002-024.

H.-C.~Cheng is supported by the Young Scholar Fellowship (Einstein Program) of the National Science and Technology Council, Taiwan (R.O.C.) under Grants No.~NSTC 112-2636-E-002-009, No.~NSTC 113-2119-M-007-006, No.~NSTC 113-2119-M-001-006, No.~NSTC 113-2124-M-002-003, 
No.~NSTC 113-2628-E-002-029,
by the Yushan Young Scholar Program of the Ministry of Education, Taiwan (R.O.C.) under Grants No.~NTU-112V1904-4 and No.~NTU-113V1904-5 and by the research project ``Pioneering Research in Forefront Quantum Computing, Learning and Engineering'' of National Taiwan University under Grant No. NTU-CC- 112L893405 and NTU-CC-113L891605. H.-C.~Cheng acknowledges the support from the “Center for Advanced Computing and Imaging in Biomedicine (NTU-113L900702)” through The Featured Areas Research Center Program within the framework of the Higher Education Sprout Project by the Ministry of Education (MOE) in Taiwan.

\bibliographystyle{unsrt}
\bibliography{sample-base}

\begin{thebibliography}{10}

\bibitem{nielsen_quantum_2010}
Michael~A. Nielsen and Isaac~L. Chuang.
\newblock {\em Quantum Computation and Quantum Information: 10th Anniversary Edition}.
\newblock Cambridge University Press, 2010.

\bibitem{khatri_principles_2020}
Sumeet Khatri and Mark~M Wilde.
\newblock Principles of quantum communication theory: A modern approach.
\newblock {\em arXiv preprint arXiv:2011.04672}, 2020.

\bibitem{chen_quantum_2015}
Chi-Yuan Chen, Guo-Jyun Zeng, Fang-jhu Lin, Yao-Hsin Chou, and Han-Chieh Chao.
\newblock Quantum cryptography and its applications over the internet.
\newblock {\em IEEE Network}, 29(5):64--69, September 2015.

\bibitem{arunachalam_survey_2017}
Srinivasan Arunachalam and Ronald De~Wolf.
\newblock Guest column: A survey of quantum learning theory.
\newblock {\em SIGACT News}, 48(2):41--67, June 2017.

\bibitem{shor_algorithms_1994}
Peter~W. Shor.
\newblock Algorithms for quantum computation: discrete logarithms and factoring.
\newblock In {\em Proc. {Annual} {Symposium} on {Foundations} of {Computer} {Science}}, pages 124--134, 1994.

\bibitem{shor_polynomial-time_1997}
Peter~W. Shor.
\newblock Polynomial-time algorithms for prime factorization and discrete logarithms on a quantum computer.
\newblock {\em SIAM Journal on Computing}, 26(5):1484--1509, 1997.

\bibitem{proos_shors_2003}
John Proos and Christof Zalka.
\newblock Shor's discrete logarithm quantum algorithm for elliptic curves.
\newblock {\em Quantum Information \& Computation}, 3(4):317–344, July 2003.

\bibitem{ibm_ibm_2021}
{IBM}.
\newblock {IBM} quantum, 2021.

\bibitem{siraichi_qubit_2018}
Marcos~Yukio Siraichi, Vinícius Fernandes~dos Santos, Caroline Collange, and Fernando Magno~Quintão Pereira.
\newblock Qubit allocation.
\newblock In {\em Proc. {International} {Symposium} on {Code} {Generation} and {Optimization}}, pages 113--125, 2018.

\bibitem{maslov_quantum_2008}
Dmitri Maslov, Sean~M Falconer, and Michele Mosca.
\newblock Quantum circuit placement.
\newblock {\em IEEE Transactions on Computer-Aided Design of Integrated Circuits and Systems}, 27(4):752--763, 2008.

\bibitem{saeedi_synthesis_2011}
Mehdi Saeedi, Robert Wille, and Rolf Drechsler.
\newblock Synthesis of quantum circuits for linear nearest neighbor architectures.
\newblock {\em Quantum Information Processing}, 10(3):355--377, 2011.

\bibitem{chakrabarti_linear_2011}
Amlan Chakrabarti, Susmita Sur-Kolay, and Ayan Chaudhury.
\newblock Linear nearest neighbor synthesis of reversible circuits by graph partitioning.
\newblock {\em arXiv preprint arXiv:1112.0564}, 2011.

\bibitem{shafaei_optimization_2013}
Alireza Shafaei, Mehdi Saeedi, and Massoud Pedram.
\newblock Optimization of quantum circuits for interaction distance in linear nearest neighbor architectures.
\newblock In {\em Proc. {ACM}/{IEEE} {Design} {Automation} {Conference} ({DAC})}, pages 1--6, 2013.

\bibitem{shafaei_qubit_2014}
Alireza Shafaei, Mehdi Saeedi, and Massoud Pedram.
\newblock Qubit placement to minimize communication overhead in {2D} quantum architectures.
\newblock In {\em Proc. {IEEE} {Asia} and {South} {Pacific} {Design} {Automation} {Conference} ({ASP}-{DAC})}, pages 495--500, 2014.

\bibitem{wille_optimal_2014}
Robert Wille, Aaron Lye, and Rolf Drechsler.
\newblock Optimal {SWAP} gate insertion for nearest neighbor quantum circuits.
\newblock In {\em Proc. {IEEE} {Asia} and {South} {Pacific} {Design} {Automation} {Conference} ({ASP}-{DAC})}, pages 489--494, 2014.

\bibitem{wille_exact_2014}
Robert Wille, Aaron Lye, and Rolf Drechsler.
\newblock Exact reordering of circuit lines for nearest neighbor quantum architectures.
\newblock {\em IEEE Transactions on Computer-Aided Design of Integrated Circuits and Systems}, 33(12):1818--1831, December 2014.

\bibitem{lye_determining_2015}
Aaron Lye, Robert Wille, and Rolf Drechsler.
\newblock Determining the minimal number of swap gates for multi-dimensional nearest neighbor quantum circuits.
\newblock In {\em Proc. {IEEE} {Asia} and {South} {Pacific} {Design} {Automation} {Conference}}, pages 178--183, 2015.

\bibitem{pedram_layout_2016}
Massoud Pedram and Alireza Shafaei.
\newblock Layout optimization for quantum circuits with linear nearest neighbor architectures.
\newblock {\em IEEE Circuits Syst. Mag.}, 16(2):62--74, 2016.

\bibitem{wille_look-ahead_2016}
Robert Wille, Oliver Keszocze, Marcel Walter, Patrick Rohrs, Anupam Chattopadhyay, and Rolf Drechsler.
\newblock Look-ahead schemes for nearest neighbor optimization of {1D} and {2D} quantum circuits.
\newblock In {\em 2016 21st {Asia} and {South} {Pacific} {Design} {Automation} {Conference} ({ASP}-{DAC})}, pages 292--297, January 2016.

\bibitem{booth_comparing_2018}
Kyle~EC Booth, Minh Do, J~Christopher Beck, Eleanor Rieffel, Davide Venturelli, and Jeremy Frank.
\newblock Comparing and integrating constraint programming and temporal planning for quantum circuit compilation.
\newblock In {\em Proc. {International} {Conference} on {Automated} {Planning} and {Scheduling}}, volume~28, pages 366--374, 2018.

\bibitem{de_almeida_cnot_2019}
Alexandre A.~A. De~Almeida, Gerhard~W. Dueck, and Alexandre C.~R. Da~Silva.
\newblock {CNOT} gate mappings to {Clifford}+{T} circuits in {IBM} architectures.
\newblock In {\em 2019 {IEEE} 49th {International} {Symposium} on {Multiple}-{Valued} {Logic} ({ISMVL})}, pages 7--12, May 2019.

\bibitem{venturelli_compiling_2018}
Davide Venturelli, Minh Do, Eleanor Rieffel, and Jeremy Frank.
\newblock Compiling quantum circuits to realistic hardware architectures using temporal planners.
\newblock {\em Quantum Science and Technology}, 3(2):025004, April 2018.

\bibitem{zulehner_efficient_2018}
Alwin Zulehner, Alexandru Paler, and Robert Wille.
\newblock An efficient methodology for mapping quantum circuits to the {IBM} {QX} architectures.
\newblock {\em IEEE Transactions on Computer-Aided Design of Integrated Circuits and Systems}, 38(7):1226--1236, 2018.

\bibitem{zulehner_compiling_2019}
Alwin Zulehner and Robert Wille.
\newblock Compiling {SU}(4) quantum circuits to {IBM} {QX} architectures.
\newblock In {\em Proceedings of the 24th {Asia} and {South} {Pacific} {Design} {Automation} {Conference}}, pages 185--190, January 2019.

\bibitem{matsuo_reducing_2019}
Atsushi Matsuo, Wakakii Hattori, and Shigeru Yamashita.
\newblock Reducing the overhead of mapping quantum circuits to {IBM} {Q} system.
\newblock In {\em 2019 {IEEE} {International} {Symposium} on {Circuits} and {Systems} ({ISCAS})}, pages 1--5, May 2019.

\bibitem{peham_optimal_2022}
Tom Peham, Lukas Burgholzer, and Robert Wille.
\newblock On optimal subarchitectures for quantum circuit mapping.
\newblock {\em ACM Transactions on Quantum Computing}, 4(4):1--20, 2023.

\bibitem{wille_mqt_2023}
Robert Wille and Lukas Burgholzer.
\newblock {MQT} {QMAP}: efficient quantum circuit mapping.
\newblock In {\em Proceedings of the 2023 {International} {Symposium} on {Physical} {Design}}, pages 198--204, 2023.

\bibitem{li_tackling_2019}
Gushu Li, Yufei Ding, and Yuan Xie.
\newblock Tackling the qubit mapping problem for {NISQ}-era quantum devices.
\newblock In {\em Proc. {International} {Conference} on {Architectural} {Support} for {Programming} {Languages} and {Operating} {Systems} (ASPLOS)}, pages 1001--1014, 2019.

\bibitem{zhu_dynamic_2020}
Pengcheng Zhu, Zhijin Guan, and Xueyun Cheng.
\newblock A dynamic look-ahead heuristic for the qubit mapping problem of {NISQ} computers.
\newblock {\em IEEE Transactions on Computer-Aided Design of Integrated Circuits and Systems}, 39(12):4721--4735, December 2020.

\bibitem{jiang_quantum_2021}
Hui Jiang, Yuxin Deng, and Ming Xu.
\newblock Quantum circuit transformation based on {Tabu} search.
\newblock {\em arXiv preprint arXiv:2011.04672}, August 2021.

\bibitem{guerreschi_two-step_2018}
Gian~Giacomo Guerreschi and Jongsoo Park.
\newblock Two-step approach to scheduling quantum circuits.
\newblock {\em Quantum Science and Technology}, 3(4):045003, 2018.

\bibitem{wille_mapping_2019}
Robert Wille, Lukas Burgholzer, and Alwin Zulehner.
\newblock Mapping quantum circuits to {IBM} {QX} architectures using the minimal number of {SWAP} and {H} operations.
\newblock In {\em Proceedings of the 56th {Annual} {Design} {Automation} {Conference} 2019}, pages 1--6, 2019.

\bibitem{tan_optimal_2020}
Bochen Tan and Jason Cong.
\newblock Optimal layout synthesis for quantum computing.
\newblock In {\em Proceedings of the 39th {International} {Conference} on {Computer}-{Aided} {Design} (ICCAD)}, pages 1--9, 2020.

\bibitem{tan_optimal_2021}
Bochen Tan and Jason Cong.
\newblock Optimal qubit mapping with simultaneous gate absorption.
\newblock In {\em 2021 {IEEE}/{ACM} {International} {Conference} {On} {Computer} {Aided} {Design} ({ICCAD})}, pages 1--8, November 2021.

\bibitem{lin_scalable_2023}
Wan-Hsuan Lin, Jason Kimko, Bochen Tan, Nikolaj Bjørner, and Jason Cong.
\newblock Scalable optimal layout synthesis for nisq quantum processors.
\newblock In {\em 2023 60th {ACM}/{IEEE} {Design} {Automation} {Conference} ({DAC})}, pages 1--6, 2023.

\bibitem{molavi_qubit_2022}
Abtin Molavi, Amanda Xu, Martin Diges, Lauren Pick, Swamit Tannu, and Aws Albarghouthi.
\newblock Qubit mapping and routing via {MaxSAT}.
\newblock In {\em 2022 55th {IEEE}/{ACM} {International} {Symposium} on {Microarchitecture} ({MICRO})}, pages 1078--1091, October 2022.

\bibitem{deng_codar_2020}
Haowei Deng, Yu~Zhang, and Quanxi Li.
\newblock Codar: A contextual duration-aware qubit mapping for various {NISQ} devices.
\newblock In {\em Proc. {ACM}/{IEEE} {Design} {Automation} {Conference} ({DAC})}, pages 1--6, 2020.

\bibitem{zhang_time-optimal_2021}
Chi Zhang, Ari~B Hayes, Longfei Qiu, Yuwei Jin, Yanhao Chen, and Eddy~Z Zhang.
\newblock Time-optimal qubit mapping.
\newblock In {\em Proc. {ACM} {International} {Conference} on {Architectural} {Support} for {Programming} {Languages} and {Operating} {Systems} (ASPLOS)}, pages 360--374, 2021.

\bibitem{lao_timing_2022}
Lingling Lao, Hans Van~Someren, Imran Ashraf, and Carmen~G. Almudever.
\newblock Timing and resource-aware mapping of quantum circuits to superconducting processors.
\newblock {\em IEEE Transactions on Computer-Aided Design of Integrated Circuits and Systems}, 41(2):359--371, February 2022.

\bibitem{brglez_neutral_1985}
Franc Brglez.
\newblock A neutral netlist of 10 combinational benchmark circuits and a target translator in {Fortran}.
\newblock In {\em Proc. {International} {Symposium} on {Circuits} and {Systems}, 1985}, 1985.

\bibitem{Lau_Qsyn_2024}
Mu-Te Lau, Chin-Yi Cheng, Cheng-Hua Lu, Chia-Hsu Chuang, Yi-Hsiang Kuo, Hsiang-Chun Yang, Chien-Tung Kuo, Hsin-Yu Chen, Chen-Ying Tung, Cheng-En Tsai, Guan-Hao Chen, Leng-Kai Lin, Ching-Huan Wang, Tzu-Hsu Wang, and Chung-Yang~Ric Huang.
\newblock Qsyn: A developer-friendly quantum circuit synthesis framework for {NISQ} era and beyond, 2024.

\bibitem{dijkstra_note_1959}
E.~W. Dijkstra.
\newblock A note on two problems in connexion with graphs.
\newblock {\em Numerische Mathematik}, 1(1):269--271, December 1959.

\bibitem{floyd_algorithm_1962}
Robert~W Floyd.
\newblock Algorithm 97: shortest path.
\newblock {\em Communications of the ACM}, 5(6):345, 1962.

\bibitem{sivarajah_tket_2021}
Seyon Sivarajah, Silas Dilkes, Alexander Cowtan, Will Simmons, Alec Edgington, and Ross Duncan.
\newblock t{\textbar}ket⟩: a retargetable compiler for {NISQ} devices.
\newblock {\em Quantum Science and Technology}, 6(1):014003, January 2021.

\bibitem{aleksandrowicz_qiskit_2019}
Gadi Aleksandrowicz, Thomas Alexander, Panagiotis Barkoutsos, Luciano Bello, Yael Ben-Haim, David Bucher, Francisco~Jose Cabrera-Hernández, Jorge Carballo-Franquis, Adrian Chen, Chun-Fu Chen, and {others}.
\newblock Qiskit: An open-source framework for quantum computing.
\newblock {\em Accessed on: Mar}, 16, 2019.

\bibitem{chamberland_topological_2020}
Christopher Chamberland, Guanyu Zhu, Theodore~J Yoder, Jared~B Hertzberg, and Andrew~W Cross.
\newblock Topological and subsystem codes on low-degree graphs with flag qubits.
\newblock {\em Physical Review X}, 10(1):011022, 2020.

\bibitem{ibm_ibm_2021-1}
{IBM}.
\newblock The {IBM} {Quantum} heavy hex lattice, 2021.

\bibitem{kim_efficient_2018}
Taewan Kim and Byung-Soo Choi.
\newblock Efficient decomposition methods for controlled-{R} n using a single ancillary qubit.
\newblock {\em Scientific reports}, 8(1):5445, 2018.

\bibitem{wille_revlib_2008}
Robert Wille, Daniel Große, Lisa Teuber, Gerhard~W Dueck, and Rolf Drechsler.
\newblock {RevLib}: An online resource for reversible functions and reversible circuits.
\newblock In {\em 38th {International} {Symposium} on {Multiple} {Valued} {Logic} (ismvl 2008)}, pages 220--225, 2008.

\bibitem{kissinger_pyzx_2020}
Aleks Kissinger and John van~de Wetering.
\newblock {PyZX}: {Large} scale automated diagrammatic reasoning.
\newblock In Bob Coecke and Matthew Leifer, editors, {\em {Proceedings} 16th {International} {Conference} on {Quantum} {Physics} and {Logic}, {Chapman} {University}, {Orange}, {CA}, {USA}., 10-14 {June} 2019}, volume 318 of {\em Electronic {Proceedings} in {Theoretical} {Computer} {Science}}, pages 229--241, 2020.

\bibitem{chen_veriqbench_2022}
Kean Chen, Wang Fang, Ji~Guan, Xin Hong, Mingyu Huang, Junyi Liu, Qisheng Wang, and Mingsheng Ying.
\newblock {VeriQBench}: A benchmark for multiple types of quantum {Circuits}.
\newblock {\em arXiv preprint arXiv:2206.10880}, 2022.

\bibitem{group_abc_2007}
VERIFICATION GROUP and {others}.
\newblock {ABC}: a system for sequential synthesis and verification, release 70930, 2007.

\bibitem{li_qasmbench_2022}
Ang Li, Samuel Stein, Sriram Krishnamoorthy, and James Ang.
\newblock Qasmbench: A low-level quantum benchmark suite for nisq evaluation and simulation.
\newblock {\em ACM Transactions on Quantum Computing}, 2022.

\bibitem{das_case_2019}
Poulami Das, Swamit~S. Tannu, Prashant~J. Nair, and Moinuddin Qureshi.
\newblock A case for multi-programming quantum computers.
\newblock In {\em Proceedings of the 52nd {Annual} {IEEE}/{ACM} {International} {Symposium} on {Microarchitecture}}, pages 291--303, October 2019.

\bibitem{tannu_not_2019}
Swamit~S. Tannu and Moinuddin~K. Qureshi.
\newblock Not all qubits are created equal: a case for variability-aware policies for {NISQ}-era quantum computers.
\newblock In {\em Proceedings of the {Twenty}-{Fourth} {International} {Conference} on {Architectural} {Support} for {Programming} {Languages} and {Operating} {Systems} (ASPLOS)}, pages 987--999, April 2019.

\bibitem{murali_noise-adaptive_2019}
Prakash Murali, Jonathan~M. Baker, Ali Javadi-Abhari, Frederic~T. Chong, and Margaret Martonosi.
\newblock Noise-adaptive compiler mappings for noisy intermediate-scale quantum computers.
\newblock In {\em Proceedings of the {Twenty}-{Fourth} {International} {Conference} on {Architectural} {Support} for {Programming} {Languages} and {Operating} {Systems} (ASPLOS)}, pages 1015--1029, April 2019.

\bibitem{li_qubits_2020}
Ze-Tong Li, Fan-Xu Meng, Zai-Chen Zhang, and Xu-Tao Yu.
\newblock Qubits’ mapping and routing for {NISQ} on variability of quantum gates.
\newblock {\em Quantum Information Processing}, 19(10):378, October 2020.

\bibitem{niu_hardware-aware_2020}
Siyuan Niu, Adrien Suau, Gabriel Staffelbach, and Aida Todri-Sanial.
\newblock A hardware-aware heuristic for the qubit mapping problem in the {NISQ} era.
\newblock {\em IEEE Transactions on Quantum Engineering}, 1:1--14, 2020.

\bibitem{das_adapt_2021}
Poulami Das, Swamit Tannu, Siddharth Dangwal, and Moinuddin Qureshi.
\newblock {ADAPT}: {Mitigating} idling errors in qubits via adaptive dynamical decoupling.
\newblock In {\em {MICRO}-54: 54th {Annual} {IEEE}/{ACM} {International} {Symposium} on {Microarchitecture}}, pages 950--962, October 2021.

\bibitem{das_jigsaw_2021}
Poulami Das, Swamit Tannu, and Moinuddin Qureshi.
\newblock {JigSaw}: {Boosting} fidelity of {NISQ} programs via measurement subsetting.
\newblock In {\em {MICRO}-54: 54th {Annual} {IEEE}/{ACM} {International} {Symposium} on {Microarchitecture}}, pages 937--949, October 2021.

\bibitem{jurcevic_demonstration_2021}
Petar Jurcevic, Ali Javadi-Abhari, Lev~S Bishop, Isaac Lauer, Daniela~F Bogorin, Markus Brink, Lauren Capelluto, Oktay Günlük, Toshinari Itoko, Naoki Kanazawa, Abhinav Kandala, George~A Keefe, Kevin Krsulich, William Landers, Eric~P Lewandowski, Douglas~T McClure, Giacomo Nannicini, Adinath Narasgond, Hasan~M Nayfeh, Emily Pritchett, Mary~Beth Rothwell, Srikanth Srinivasan, Neereja Sundaresan, Cindy Wang, Ken~X Wei, Christopher~J Wood, Jeng-Bang Yau, Eric~J Zhang, Oliver~E Dial, Jerry~M Chow, and Jay~M Gambetta.
\newblock Demonstration of quantum volume 64 on a superconducting quantum computing system.
\newblock {\em Quantum Science and Technology}, 6(2):025020, April 2021.

\bibitem{mundada_suppression_2019}
Pranav Mundada, Gengyan Zhang, Thomas Hazard, and Andrew Houck.
\newblock Suppression of qubit crosstalk in a tunable coupling superconducting circuit.
\newblock {\em Phys. Rev. Applied}, 12(5):054023, November 2019.

\bibitem{murali_software_2020}
Prakash Murali, David~C. Mckay, Margaret Martonosi, and Ali Javadi-Abhari.
\newblock Software mitigation of crosstalk on noisy intermediate-scale quantum computers.
\newblock In {\em Proceedings of the {Twenty}-{Fifth} {International} {Conference} on {Architectural} {Support} for {Programming} {Languages} and {Operating} {Systems}}, pages 1001--1016, March 2020.

\bibitem{ding_systematic_2020}
Yongshan Ding, Pranav Gokhale, Sophia~Fuhui Lin, Richard Rines, Thomas Propson, and Frederic~T. Chong.
\newblock Systematic crosstalk mitigation for superconducting qubits via frequency-aware compilation.
\newblock In {\em 2020 53rd {Annual} {IEEE}/{ACM} {International} {Symposium} on {Microarchitecture} ({MICRO})}, pages 201--214, October 2020.

\bibitem{ohkura_simultaneous_2022}
Yasuhiro Ohkura, Takahiko Satoh, and Rodney Van~Meter.
\newblock Simultaneous execution of quantum circuits on current and near-future {NISQ} systems.
\newblock {\em IEEE Transactions on Quantum Engineering}, 3:1--10, 2022.

\bibitem{hillmich_exploiting_2021}
Stefan Hillmich, Alwin Zulehner, and Robert Wille.
\newblock Exploiting quantum teleportation in quantum circuit mapping.
\newblock In {\em Proceedings of the 26th {Asia} and {South} {Pacific} {Design} {Automation} {Conference}}, pages 792--797, January 2021.

\bibitem{pozzi_using_2022}
Matteo~G. Pozzi, Steven~J. Herbert, Akash Sengupta, and Robert~D. Mullins.
\newblock Using reinforcement learning to perform qubit routing in quantum compilers.
\newblock {\em ACM Transactions on Quantum Computing}, 3(2):1--25, June 2022.

\bibitem{childs_circuit_2019}
Andrew~M. Childs, Eddie Schoute, and Cem~M. Unsal.
\newblock Circuit transformations for quantum architectures.
\newblock In {\em Conference on the Theory of Quantum Computation, Communication and Cryptography (TQC)}, 2019.

\bibitem{itoko_quantum_2019}
Toshinari Itoko, Rudy Raymond, Takashi Imamichi, Atsushi Matsuo, and Andrew~W. Cross.
\newblock Quantum circuit compilers using gate commutation rules.
\newblock In {\em Proceedings of the 24th {Asia} and {South} {Pacific} {Design} {Automation} {Conference}}, pages 191--196, January 2019.

\bibitem{itoko_optimization_2020}
Toshinari Itoko, Rudy Raymond, Takashi Imamichi, and Atsushi Matsuo.
\newblock Optimization of quantum circuit mapping using gate transformation and commutation.
\newblock {\em Integration}, 70:43--50, January 2020.

\bibitem{liu_not_2022}
Ji~Liu, Peiyi Li, and Huiyang Zhou.
\newblock Not all {SWAPs} have the same cost: a case for optimization-aware qubit routing.
\newblock In {\em 2022 {IEEE} {International} {Symposium} on {High}-{Performance} {Computer} {Architecture} ({HPCA})}, pages 709--725, April 2022.

\end{thebibliography}

\end{document}